\documentclass[pre,showpacs,twocolumn,amsfonts,amsmath]{revtex4}
\usepackage[final]{graphicx}

\begin{document}
\title{Tagged-particle dynamics in a hard-sphere system:
       mode-coupling theory analysis}
\date{\today}
\def\edin{%
  \affiliation{University of Edinburgh, School of Physics,
  JCMB The Kings Buildings, Mayfield Road, Edinburgh EH9 3JZ, Scotland}}
\def\konstanz{%
  \affiliation{Fachbereich Physik, Universit\"at Konstanz,
  Universit\"atsstra{\ss}e 10, 78457 Konstanz, Germany}}
\author{Th.~Voigtmann}\edin\konstanz
\author{A.~M.~Puertas}
\affiliation{Group of Complex Fluid Physics, Department of Applied Physics,
             University of Almeria, 04120 Almeria, Spain}
\author{M.~Fuchs}\konstanz

\begin{abstract}
The predictions of the mode-coupling theory of the glass
transition (MCT) for the tagged-particle density-correlation functions and
the mean-squared displacement curves are compared quantitatively and in detail
to results from
Newtonian- and Brownian-dynamics simulations of a polydisperse
quasi-hard-sphere system close to the glass transition.
After correcting for a 17\% error in the dynamical length scale and for
a smaller error in the transition density, good agreement is found over
a wide range of wave numbers and
up to five orders of magnitude in time. Deviations are found at the
highest densities studied, and for small wave vectors and the mean-squared
displacement.
Possible error sources not related to MCT are discussed in detail, thereby
identifying more clearly the issues arising from the MCT approximation
itself. The range of applicability of MCT for the different types of
short-time dynamics is established through asymptotic analyses of the
relaxation curves, examining the wave-number and density-dependent
characteristic parameters. Approximations made in the description of the
equilibrium static structure are shown to have a remarkable effect on the
predicted numerical value for the glass-transition density. Effects of
small polydispersity are also investigated, and shown to be negligible.
\end{abstract}

\maketitle

\section{Introduction}\label{intro}

Understanding the slow dynamical processes that occur when one cools
or compresses a liquid is a great challenge of condensed matter physics.
In particular in the time window accessible to scattering experiments or
molecular-dynamics (MD) computer simulations, one observes in equilibrium
a precursor of the liquid-glass transition that is commonly termed
structural relaxation. From these experiments, a large amount of detailed
information about the equilibrium fluctuations in such systems
is available \cite{Goetze1999}.

Many of the recent experiments on structural relaxation were stimulated
by the mode-coupling theory of the glass transition (MCT). This theory
attempts to provide a first-principles description of the slow structural
relaxation processes, requiring as input the (averaged) equilibrium static
structure of the system under study. Unfortunately, for many commonly
studied glass formers, the latter is not available to the extent required.
Thus comparisons of MCT with experiment usually proceed by referring to
asymptotic predictions or schematic simplifications of the theory that can be
evaluated without restriction to a specific system, and by fitting the
remaining parameters of the theory. One has to be careful when interpreting
these results, since it is known that most experimental data is hardly inside
the regime of applicability of the asymptotic formulas \cite{Franosch1997,
Goetze2000b}. Still, this way, many
studies of the predicted MCT scenario have been performed
(see Ref.~\cite{Goetze1999} for a review).

Having established the general scenario, important questions arising
are what are its ranges of validity, and what is the effect of
the approximations made in the course of deriving the theory. These
questions can be addressed by comparing the `full' solutions of the
theory to experimental results for one and the same system for which the
static structure is known in detail.
While work has been done along this direction to various degrees of detail
recently, a coherent picture for a single prototypical system has not yet
emerged.
This paper aims towards filling this gap by providing a detailed comparison
of computer-simulation results for a system of quasi-hard spheres with the
corresponding `full' MCT solutions, to establish the performance of MCT in
describing the dynamics of a prototypical glass-forming system as a fully
microscopic theory.

Such first-principles comparisons have become possible with the appearance
of powerful MD simulations for simple model systems.
Simulation data has been used to successfully test the MCT predictions
for the frozen glassy structure (the long-time limit of the dynamical
correlation functions) for a mixture of Lennard-Jones particles
\cite{Nauroth1997}, a model silica melt \cite{Sciortino2001}, and a
computer model of the molecular glass former Ortho-terphenyl \cite{Chong2003}.
In these cases, the equilibrium-structure input to MCT was determined from
the simulations themselves. The dynamical information has not been compared
to MCT in these cases. This comparison has been tackled for the
Lennard-Jones mixture \cite{Kob2002} and for two binary hard-sphere
mixtures \cite{Foffi2004}, but there the discussion had to be restricted
to the slowest decay process, while qualitative deviations from MCT at
intermediate and short times could not be resolved. This is in contrast to
a full-MCT analysis of experimental light-scattering data from a quasi-binary
hard-sphere like colloidal mixture \cite{Voigtmann2003}, where agreement over
the full accessible time window was found as far as MCT was concerned,
including short and intermediate times. It is unclear to what extent the
different system types and the different forms of short-time dynamics
between the MD simulations and the colloidal system give rise to the
differing results.
Thus it seems appropriate to perform this comparison for an even more
fundamental, paradigmatic glass former, viz.: the hard-sphere system (HSS).

Simulations for this system close to the glass transition have been performed
by Doliwa and Heuer \cite{Doliwa1998,Doliwa2000} using a Monte Carlo procedure
and a slight polydispersity.
There, an emphasis was put on the analysis of cooperative motion on the
single-particle level, and no quantitative connection to MCT was reported.
Instead we focus on the analysis of the self-intermediate scattering
functions, which can be directly compared to theory and experiment.
We chose to perform molecular dynamics (MD) simulations instead of MC,
in order to be able to also study the influence of different realistic
types of short-time dynamics, i.e.\ `atomistic' Newtonian dynamics (ND)
and `colloidal' Brownian dynamics (BD). Such a study has been performed
earlier for the Lennard-Jones mixture mentioned above \cite{Gleim1998},
however no full-MCT analysis was included there.

For an observation of the equilibrium glassy dynamics, it is, in general,
necessary to avoid crystallization by some means. For the HSS, this can
be accomplished by introducing a small amount of polydispersity that
drastically reduces crystallization rates \cite{Williams2001}. This is
inherently the case in studies of colloidal suspensions. In the
MD simulation, we are able to fully control the distribution
of particle radii in the system.
In colloidal suspensions, solvent-mediated hydrodynamic interactions
(HI) are inevitable. It is an as yet not settled question
to what extent HI influence the dynamics at high densities.
In the present simulations, HI are
not present. Thus our study also serves to complement previous
analyses of colloidal hard-sphere suspensions with asymptotic formulas
\cite{Megen1993,Megen1994b},
demonstrating that HI are not an important ingredient for a quantitative
description of structural relaxation.

The paper is organized as follows. First, we introduce in Sec.~\ref{details}
the relevant quantities for the discussion. An investigation of some asymptotic
properties of the simulation data is performed in Sec.~\ref{data-analysis},
whereas Sec.~\ref{mct-analysis} is devoted to a comparison of the time-dependent
data with MCT results for the one-component HSS.
In Sec.~\ref{polydisp}, the effects of polydispersity will be discussed
within the framework of this MCT analysis. We summarize our findings
in the conclusions, Sec.~\ref{conclusion}.

\section{Simulation and Theory Details}\label{details}

\subsection{Molecular Dynamics Simulations}

We perform standard molecular-dynamics simulations of $N=1000$ particles
in the canonical ensemble
in a polydisperse system of quasi-hard spheres. The core-core repulsion
between particles at a distance $r$ is given by
\begin{equation}
V_c(r)=k_{\text{B}}T\left( \frac{r}{d_{12}} \right)^{-36}\,,
\end{equation}
where $d_{12}$ is the center-to-center distance, $d_{12}=(d_1+d_2)/2$,
with $d_1$ and $d_2$ the diameters of the particles. This potential is
tailored to be a continuous approximation to the hard-sphere potential
considered in the theoretical part of the work, as this facilitates the
simulation of Brownian dynamics.
The control parameters of this soft-sphere system
are the number density $\varrho$ and temperature $T$; they appear however
only as a single effective coupling parameter, $\Gamma=\varrho T^{-12}$
\cite{Hansen1986}. In the simulation, $\Gamma$ is varied by changing
the density and keeping the temperature fixed. In the
following, we denote the number density in terms of a packing (or volume)
fraction, which for a monodisperse system reads $\varphi=(\pi/6)\varrho d^3$.
To avoid crystallization, the diameters of the particles in the simulation
are distributed according to a flat distribution centered around $d$ and a
half-width of $\delta/2=0.1d$. Thus the volume fraction reads
$\varphi=(\pi/6)d^3\left[1+\delta^2\right]\varrho$. 

Note that, due to polydispersity and finite-size effects, it is not trivial
to ensure that the volume fraction remains constant among different
runs, i.e.\ different realizations of the polydispersity distribution.
If one randomly draws $N$ particles with radii according to the polydispersity
distribution at a fixed number density, the resulting packing fraction will
vary from run to run, by up to about $1\%$ in the cases we have investigated.
This is not acceptable, since the slow dynamics to be discussed depends
sensitively on the packing fraction. In order to eliminate such fluctuations
of $\varphi$, we instead choose a fixed realization of the radius distribution
($1000$ equally spaced radii from $0.9$ to $1.1$), and randomly assign each
radius to one of the particles in the initial configuration.

Both Newtonian and Brownian dynamics simulations were performed, to analyze
the effect of the microscopic dynamics on the structural relaxation.
Newtonian dynamics (ND) was simulated
by integrating the Newton's equations of motion in the canonical ensemble
at constant volume. In Brownian dynamics (BD), or more precisely, strongly
damped Newtonian dynamics,
each particle experiences a Gaussian distributed white noise force with zero
mean, $\vec{\eta}(t)$, and a damping force proportional
to the velocity, $\gamma \dot{\vec{r}}$, apart from the deterministic forces
from the interactions. Hence the equation of motion for particle $j$ is
\begin{equation}\label{bdeq}
m \ddot{\vec{r}}_j - \sum_i \vec{F}_{ij}
  = -\gamma \dot{\vec{r}}_j + \vec{\eta}_j(t)\,,
\end{equation}
where $\gamma$ is a damping constant. The stochastic and friction forces
fulfill the fluctuation-dissipation theorem, $\langle \vec{\eta}_i(t)
\vec{\eta}_j(t') \rangle=6 k_{\text{B}}T \gamma \delta(t-t') \delta_{ij}$.
The value of $\gamma$ was set to $(30/\sqrt{3})\,kT/(d v_{\text{th}})$;
this `overdamped limit' ensures that
the results presented here no longer show a dependence on the value $\gamma$.
Such a form of the dynamics has been introduced in the study of glassy
relaxation by Gleim \textit{et~al.}\ \cite{Gleim1998}.
Let us note that the short-time dynamics visible in the correlators and in the
mean-squared displacement still is not strictly diffusive, but rather strongly
damped ballistic. Since it is not our aim to investigate the very short-time
dynamical features of the simulations, this will not be discussed in the
following.

Equilibration runs were done with ND in all cases, since the damping not only
introduces a change in the overall time scale but also slows down the
equilibration process. Lengths are measured in units of the diameter, the
unit of time is fixed setting the thermal velocity to
$v_{\text{th}}=\sqrt{k_{\text{B}}T/m}=1/\sqrt{3}$, and the temperature is
fixed to $k_{\text{B}}T=1/3$.
In ND, the equations of motion were integrated using the velocity-Verlet
algorithm \cite{FrenkelSmit},
with a time step $\delta t=0.0025$. The thermostat was
applied by rescaling the particle velocity to ensure constant temperature every
$n_t$ time steps. For well equilibrated samples, no effect of $n_t$ was
detected. The equations for Brownian motion were integrated following a Heun
algorithm \cite{paul95} with a time step of $\delta t=0.0005$. In this case,
no external thermostat was used, since the samples were already equilibrated
when running BD simulations. 

The orientational order parameter $Q_6$ \cite{steinhardt83,wolde96} was used
to check that the system was not crystalline. For amorphous liquid-like
structures,
$Q_6$ is close to zero, whereas it takes a finite value for an ordered phase.
Even though the $10\%$ polydispersity is high enough to prevent
crystallization in most cases, some samples at the highest volume fractions
studied did show a tendency to crystallize.
Those have been excluded from the analysis.

The volume fractions investigated in this work are $\varphi=0.50$,
$0.53$, $0.55$, $0.57$, $0.58$, $0.585$, and $0.59$. At each volume fraction,
we extracted as statistical information on the slow dynamics the
self part of the intermediate scattering function for several wave vectors
$\vec q$, $\phi^s(q,t)=\langle\exp[-i\vec q(\vec r_s(t)-\vec r_s(0))]\rangle$,
formed with the Fourier-transformed fluctuating density of a single
`tagged' particle at position $\vec r_s(t)$. Here, angular brackets
$\langle\cdot\rangle$ denote canonical averaging. A related quantity which
we also extracted from the simulations is the mean-squared displacement (MSD)
of a tagged particle, $\delta r^2(t)=\langle(\vec r_s(t)-\vec r_s(0))^2\rangle$.
The correlators and the MSD were averaged over typically $50$ runs, except
for the BD simulations at $\varphi=0.585$ and $0.59$, where $20$ runs have
been performed originally. For $\varphi=0.59$ we have also performed
additional runs for both ND and BD in order to investigate some phenomena
found there, see Sec.~\ref{alpha-analysis} below.

\subsection{Mode-Coupling Theory}\label{mctdetails}

In a system of $N$ structureless classical particles, i.e.\ without
any internal degrees of freedom, the statistical information on
the structural dynamics is encoded in the density correlation function,
$\phi(q,t)=\langle\varrho(\vec q,t)^*\varrho(\vec q)\rangle$, formed with
the fluctuating number densities $\varrho(\vec q,t)=\sum_{j=1}^N\exp(i\vec q
\cdot\vec r_j(t))/\sqrt N$ for wave vector $\vec q$.
$\phi(q,t)$ is a real
function that depends on $\vec q$ only through $q=|\vec q|$, since it is the
Fourier transform of a real, translational-invariant and isotropic function.
The dynamics given by $\phi(q,t)$ is probed by the mean-squared displacement,
$\delta r^2(t)$, and the self-part of the intermediate scattering function
(also called tagged-particle correlation function), $\phi^s(q,t)$,
extracted from our simulations. Note that the latter is linked to the
MSD in the limit $q\to0$, via $\phi^s(q,t)=1-(1/6)q^2\delta r^2(t)
+{\mathcal O}(q^4)$.

The mode-coupling theory of the glass transition (MCT)
\cite{Goetze1991b} builds upon an
exact equation of motion for the density autocorrelation function
$\phi(q,t)$,
\begin{subequations}\label{mcteq}
\begin{multline}
  \frac1{\Omega(q)^2}\partial_t^2\phi(q,t)+\phi(q,t)\\
  +\int_0^t m(q,t-t')\partial_{t'}\phi(q,t')\,dt'=0\,.
\end{multline}
Here, $\Omega(q)^2=q^2v_{\text{th}}^2/S(q)$ is a characteristic squared
frequency of the short-time motion. The equation of motion
is supplemented by the initial conditions $\phi(q,t=0)=1$ and
$\partial_t\phi(q,t=0)=0$. All many-body interaction
effects are contained in the memory kernel $m(q,t)$, the description of
which is the aim of the MCT approximations. One splits off from this
kernel a mode-coupling contribution $m^{\text{MCT}}(q,t)$, while the
remainder is assumed to describe regular liquid-state dynamics. Let us
approximate this latter part by an instantaneous contribution,
\begin{equation}\label{mctmemreg}
  m(q,t)\approx(\nu(q)/\Omega(q)^2)\delta(t)+m^{\text{MCT}}(q,t)\,.
\end{equation}
The damping term $\nu(q)$ is chosen as $\nu=(30/\sqrt{3})\,v_{\text{th}}/d$
independent of $q$; a choice that ensures the short-time expansion of
$\phi(q,t)$ in the overdamped limit
to match that one of the simulation, cf.\ Eq.~\eqref{bdeq}: one
gets
$\phi(q,t)=1-(q^2/S(q))(k_{\text{B}}T/\gamma)t+{\mathcal O}(t^2)
=1-(\Omega(q)^2/\nu)t+{\mathcal O}(t^2)$.
Note that the $q$-independent choice of $\nu$ destroys momentum conservation
for the hard-sphere particles; this is appropriate for a model of a colloidal
system.

The MCT contribution to the memory kernel is given by
$m^{\text{MCT}}(q,t)={\mathcal F}_q[\phi(t)]$,
where
\begin{equation}\label{memory}
  {\mathcal F}[\hat f]=
  \frac\varrho{2q^4}\int\frac{d^3k}{(2\pi)^3}S(q)S(k)S(p)V(\vec q,\vec k,\vec p)
  \hat f(k)\hat f(p)
\end{equation}
and the abbreviation $\vec p=\vec q-\vec k$ has been used. The vertices
$V(\vec q,\vec k,\vec p)$ are the coupling constants of the theory, through
which all crucial control-parameter dependence enters. They are given entirely
in terms of static two- and three-point correlation functions describing
the equilibrium structure of the system's liquid state. The latter are
approximated using a convolution approximation, so that the vertex reads
\begin{equation}\label{mctv}
  V(\vec q,\vec k,\vec p)
  = \left[(\vec q\vec k)c(k)+(\vec q\vec p)c(p)\right]^2\,.
\end{equation}
Here, $c(q)$ is the direct correlation function (DCF) connected to the
static structure factor by $S(q)=1/(1-\varrho c(q))$.
\end{subequations}

The long-time limit of the correlation functions, $f(q)=\lim_{t\to\infty}
\phi(q,t)$, is used to discriminate between liquid and glassy states. In the
liquid, $f(q)\equiv0$, while the glass is characterized by some $f(q)\neq0$.
From Eqs.~\eqref{mcteq}, one finds $f(q)$ as the largest real and positive
solution of the implicit equations \cite{Goetze1995b}
\begin{equation}\label{mctfeq}
  \frac{f(q)}{1-f(q)}={\mathcal F}_q[f]\,.
\end{equation}
In particular, there exist critical points in the control-parameter space,
identified as ideal glass transition points, where a new permissible
solution of Eq.~\eqref{mctfeq} appears.
Typically, $f(q)$ jumps discontinuously from zero to nonzero values there.
Close to such a critical point on the liquid side, the correlation functions
exhibit a two-step relaxation scenario, composed by a relaxation towards a
plateau value, and by a later relaxation from this plateau value to zero
termed $\alpha$ relaxation. On approaching the transition, the characteristic
time scale for the $\alpha$ relaxation diverges, and an increasingly large
window opens where the correlation functions stay close to their plateau.
The plateau values on the liquid side are in leading order given by the
critical solutions of Eq.~\eqref{mctfeq}, $f^c(q)$, i.e.\ by the maximal
solutions of Eq.~\eqref{mctfeq} evaluated at a critical point.
The time window for which $\phi(q,t)$ is close to $f^c(q)$ is called the
$\beta$-relaxation regime, and is the object of asymptotic predictions of
MCT \cite{Goetze1991b,Franosch1997,Fuchs1998}. These include scaling laws
for the correlators, whose power-law exponents $a$ and $b$, called the
critical and the von~Schweidler exponent, are given by an exponent parameter
$\lambda$. The latter is calculated within MCT and depends on the static
equilibrium structure of the system. We will test some of the predictions
connected with $\beta$ relaxation in Sec.~\ref{betarelaxation}.

Let us also recollect the MCT equations of motion for the tagged-particle
correlation function $\phi^s(q,t)$ of a tagged particle that is of the
same species as the host fluid, since this will be the quantity we
shall analyze below. For it, an expression similar to that of
Eqs.~\eqref{mcteq} holds,
\begin{subequations}\label{mctseq}
\begin{multline}
  \frac1{\Omega^s(q)^2}\partial_t^2\phi^s(q,t)+\phi^s(q,t)\\
  +\int_0^t m^s(q,t-t')\partial_{t'}\phi^s(q,t')\,dt'=0\,,
\end{multline}
where we have $\Omega^s(q)^2=q^2v_{\text{th}}^2$.
The tagged-particle memory kernel is given in MCT approximation by
$m^s(q,t)\approx(\nu_s(q)/Omega^s(q)^2)\delta(t)
+{\mathcal F}^s[\phi^s(t),\phi(t)]$, with
\begin{equation}\label{memorytagged}
  {\mathcal F}^s[\hat f^s,\hat f]
  =\frac1{q^4}\int\frac{d^3k}{(2\pi)^3}V^s(\vec q,\vec k)f(k)f^s(p)\,,
\end{equation}
and with vertices
\begin{equation}\label{mctvs}
  V^s(\vec q,\vec k)
  = (\vec q\vec k)^2c(k)^2\,,
\end{equation}
where we set $\nu_s(q)\equiv\nu$ in the following.
\end{subequations}
The qualitative features of $\phi^s(q,t)$ close to an ideal glass transition
are the same as those of $\phi(q,t)$, as long as it couples strongly enough
to via Eq.~\eqref{memorytagged}. In this generic case, also $\phi^s(q,t)$
develops a two-step
relaxation pattern, with plateaus given by the critical solution $f^{s,c}(q)$
of the tagged-particle analog of Eq.~\eqref{mctfeq},
\begin{equation}\label{mctfseq}
  \frac{f^s(q)}{1-f^s(q)}={\mathcal F}^s_q[\hat f,\hat f^s]\,.
\end{equation}

The mean-squared displacement (MSD) $\delta r^2(t)$ can be calculated
from the $q\to0$ limit of the tagged-particle correlation function.
One gets
\begin{equation}\label{msdequation}
  \partial_t\delta r^2(t)+v_{\text{th}}^2\int_0^t m_0^s(t-t')
  \delta r^2(t')\,dt'=6v_{\text{th}}^2t\,,
\end{equation}
where we have set $m_0^s(t)=\lim_{q\to0}q^2m^s(q,t)$.

Eqs.~\eqref{mcteq} can be solved numerically for the functions $\phi(q,t)$,
once the vertices $V(q,k,p)$ have been calculated from liquid-state theory.
To this end, the wave vectors are discretized on a regular grid of $M$
wave numbers with spacing $\Delta_q$: $qd=i\Delta_q+q_0$. We have used
$M=300$, $\Delta_q=0.4/3$, and $q_0=0.2/3$, implying a cutoff wave vector
$q^*d=39.93$. This discretization is enough to ensure that the
long-time part of the dynamics does not show significant numerical
artifacts \cite{Franosch1997} and has been used before in the discussion of
MCT results for the HSS \cite{Goetze2000}. Once the $\phi(q,t)$ have been
determined, a similar numerical scheme allows to evaluate Eqs.~\eqref{mctseq}
for the $\phi^s(q,t)$, and from this, one gets $\delta r^2(t)$ from
Eq.~\eqref{msdequation}.

For the solution of Eqs.~\eqref{mctfeq}, a straightforward iteration
scheme guarantees a numerically stable determination of the correct
solutions $f(q)=\phi(q,t\to\infty)$ \cite{Goetze1995b} and, once the $f(q)$ are
calculated, of $f^s(q)=\phi^s(q,t\to\infty)$. From the distinction between
states with $f(q)\neq0$ and $f(q)\equiv0$, the critical point $\varphi^c$
can be found by iteration in $\varphi$. For the solution of these
equations, we have used a discretization with $M=100$, $\Delta_q=0.4$,
and $q_0=0.2$. This is sufficient to ensure that errors in the $f(q)$,
$f^s(q)$, and $\varphi^c$ resulting from the
different discretizations used are small.

A few results shall also be discussed concerning the polydispersity-induced
effects. MCT for continuous polydispersity distributions is not available,
but we try to estimate the influence of the polydispersity by calculating
MCT results for $S$-component mixtures with the species' diameters chosen to
mimic the simulated polydisperse distribution. We have used an $S=3$ model
with diameters $d_\alpha\in\{1-w,1,1+w\}$, and $\varrho_\alpha=\varrho/3$, where
$\alpha$ labels the species of the mixture, and $\varrho_\alpha$
is the partial number density of each species. Here, we set $w=1/\sqrt{200}$
in order to match the second moment of the discrete distribution to that of
the one used in the simulation. We have also calculated
results for an $S=5$ model, with $d_\alpha\in\{0.9,0.95,1.0,1.05,1.1\}$,
and $\varrho_\alpha=\varrho/5$, chosen to contain particles within the same
size range as in the simulation. The MCT equations, Eqs.~\eqref{mcteq},
generalize to mixtures in an obvious way, leading to equations of motion
for the matrix of partial density correlators, $\Phi_{\alpha\beta}(q,t)$
\cite{Goetze1987b,Goetze2003}.
Similar to Eqs.~\eqref{mctseq}, the correlators for a
tagged particle of either one of the species, $\phi^s_\alpha(q,t)$, are
calculated, together with their long-time limits, $f^s_\alpha(q)$. We
can now define `averaged' tagged-particle quantities as
\begin{equation}\label{ncompavg}
  f_{\text{pd}}^s(q)=\frac1S\sum_{\alpha=1}^Sf_\alpha^s(q)\,,
\end{equation}
and similarly for $\phi^s_{\text{pd}}(q,t)$. These quantities are analogous
to the quantities extracted from the polydisperse MD simulations.

To calculate results from the MCT equations, we require as the only input
expressions for the direct correlation function $c(q)$ entering the vertices,
Eqs.~\eqref{mctv} and \eqref{mctvs}.
For the multi-component
analog of these expressions, one requires knowledge of the full matrix of
direct correlation functions, $c_{\alpha\beta}(q)$. The DCF could be either
determined from simulations, or taken from well-known results of
liquid structure theory. For hard spheres, the Percus-Yevick (PY) closure
to the Ornstein-Zernike equation provides a fairly accurate parameter-free
description \cite{Hansen1986}. Using the PY-DCF as input to MCT, we thus
obtain results for the dynamics of the HSS that are independent on any
empirical parameters or any (usually not readily available) simulation input.
These predictions are the best we can currently achieve from within MCT
as a completely parameter-free theory. Furthermore, the wealth
of asymptotic predictions of MCT has been worked out in great detail for
this model \cite{Franosch1997,Fuchs1998}.
Note that the PY approximation to the DCF itself introduces errors that are
independent from those introduced by the MCT approximation. It has been
pointed out recently that these PY-induced errors can be quite pronounced
in the MCT-calculated quantities, even if they appear small at the $S(q)$
level \cite{Foffi2004}. To disentangle these two error sources, we have also
performed some calculations within MCT with $S(q)$ obtained from our
simulation, as will be discussed below.

\section{Data Analysis}\label{data-analysis}

\begin{figure}
\includegraphics[width=.75\linewidth]{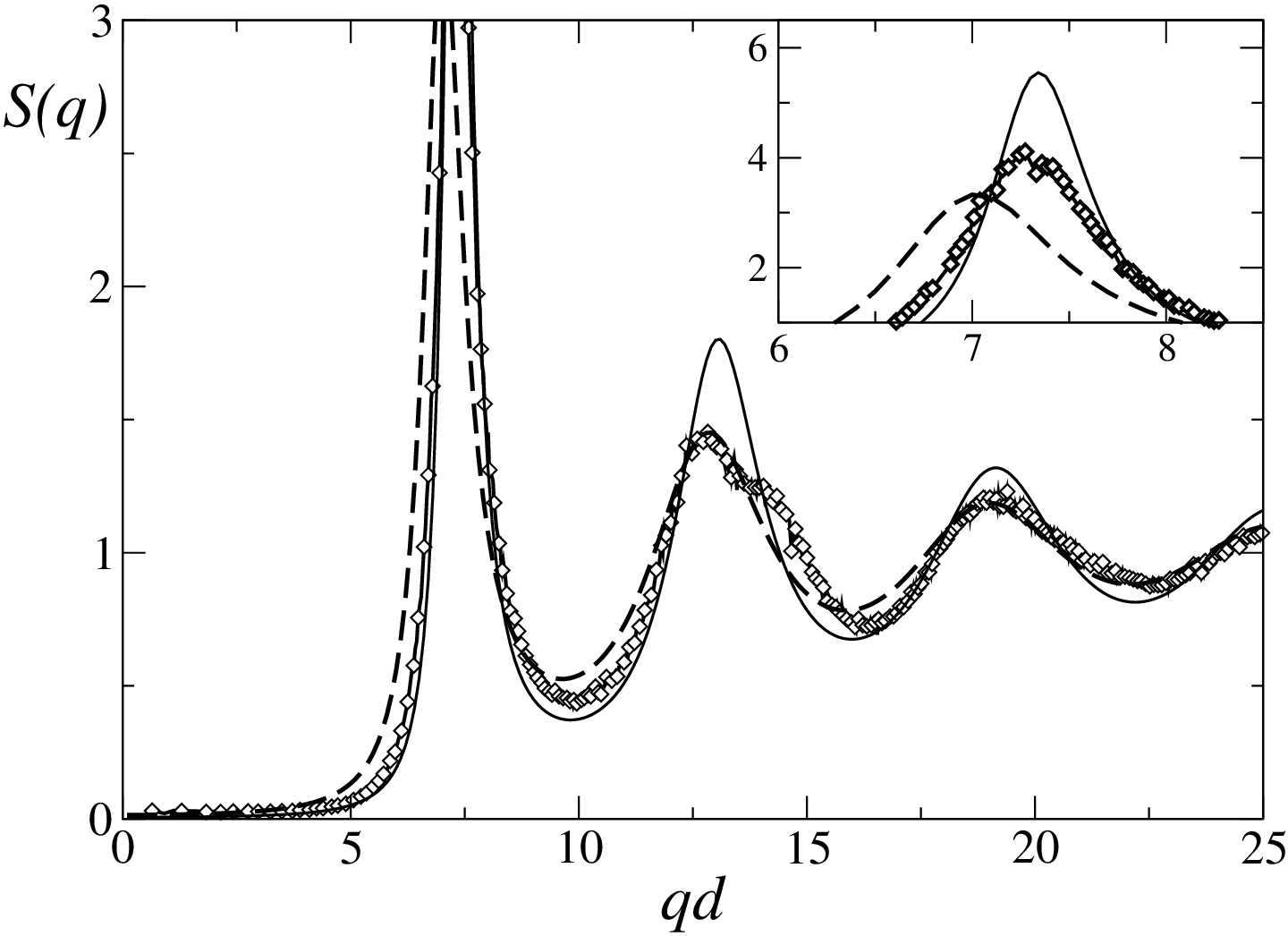}
\caption{\label{sq-comp}
  Comparison of the static structure factor $S(q)$ for the simulated
  polydisperse soft-sphere system at $\varphi=0.58$ (symbols) with the
  Percus-Yevick approximation to the hard-sphere $S(q)$ at the same
  density (solid line). The dashed line shows the Percus-Yevick result
  for $\varphi=0.505$. The region around the first diffraction peak
  is enlarged in the inset.
}
\end{figure}

Let us start the discussion of the data by a comparison of the structure
factor $S(q)$ obtained from the simulation with the PY approximation, since
this is the crucial input to all MCT calculations below.
Fig.~\ref{sq-comp} shows this quantity for $\varphi=0.58$. While PY
reproduces the oscillation period in $S(q)$, i.e.\ the typical length
scale, correctly, it overestimates the peak heights, i.e.\ the strength
of ordering in the system \cite{Hansen1986}.
Since the strength of the coupling constants in MCT is directly connected
to the peak heights in $S(q)$, the MCT calculation based on the PY $S(q)$
will overestimate the tendency to glass formation. One can try to adjust
the peak heights by setting a lower packing fraction in the PY calculation.
This is demonstrated by the dashed line in Fig.~\ref{sq-comp}, where
$\varphi=0.505$ has been taken. This value has in fact been determined by
the MCT fits presented in Sec.~\ref{mct-analysis}, and is chosen such
that the final relaxation time in the MCT calculations at that density
matches the one of the simulations at $\varphi=0.58$. As Fig.~\ref{sq-comp}
demonstrates, this introduces a small error in the oscillation period
in $S(q)$.

\begin{figure}
\includegraphics[width=.75\linewidth]{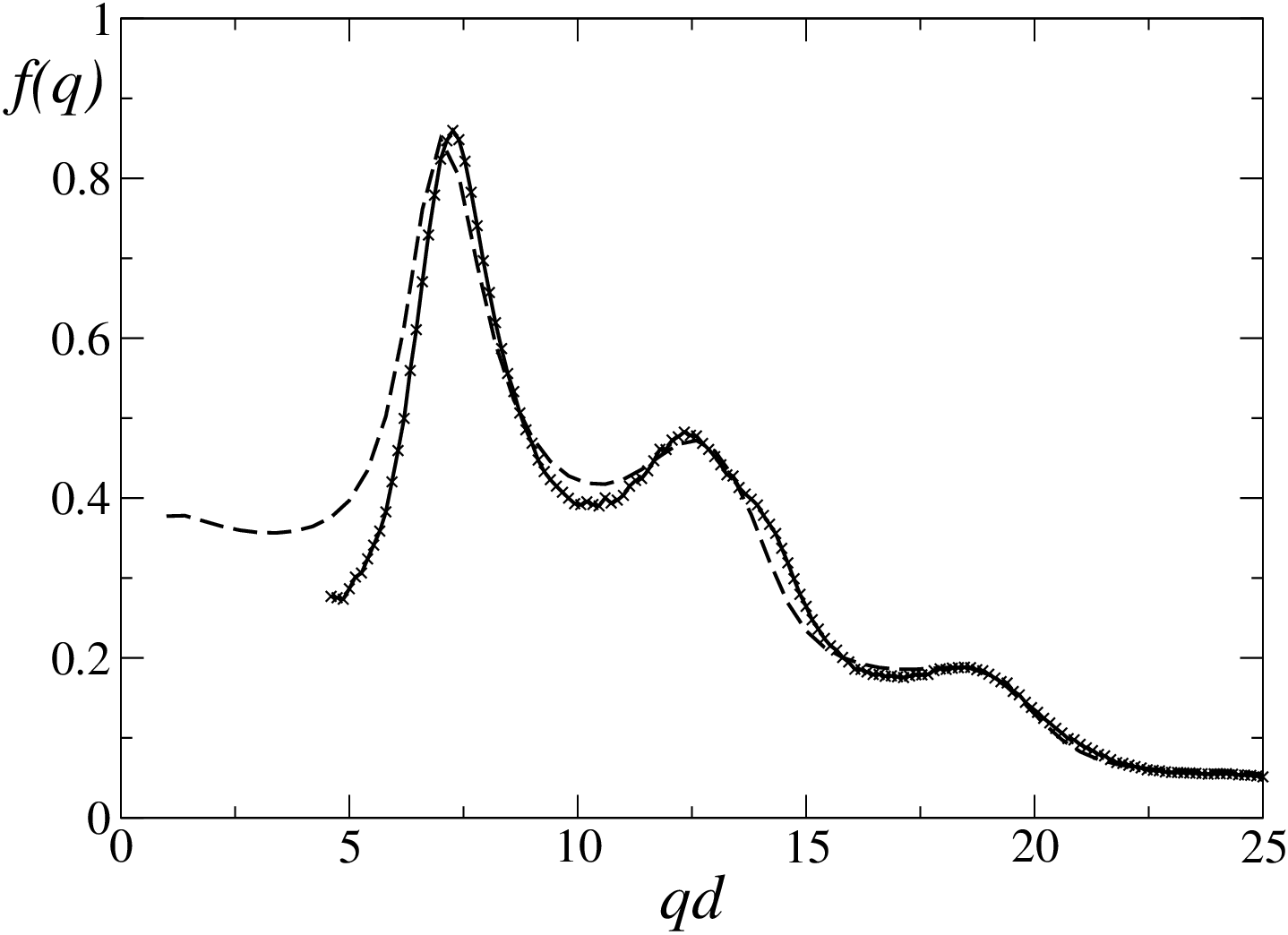}
\caption{\label{simsq-fc}
  MCT results for the critical nonergodicity parameters $f^c(q)$, using
  as input the static structure factor $S(q)$ from Percus-Yevick theory
  for hard spheres (dashed line). The crosses connected by the solid line
  show the results using $S(q)$ as obtained from the polydisperse
  nearly-hard-sphere simulation.
}
\end{figure}

It is well known that MCT, based on the PY structure factor for hard
spheres, underestimates the glass-transition packing fraction of that system.
One gets $\varphi^c_{\text{MCT}}\approx0.516$ \cite{Franosch1997},
instead of the value reported from experiments on colloidal hard-sphere-like
suspensions, $\varphi^c\approx0.58$
\cite{Megen1993}. In order to determine to which extent such an underestimation
can stem from deviations of PY from the simulated $S(q)$, which are
visible in Fig.~\ref{sq-comp}, we have calculated MCT results for
$\varphi^c_{\text{MCT}}$ and the critical plateau values $f^c(q)$ both using the
PY approximation and using our simulation results for $S(q)$ as input to
the theory.
We have evaluated the structure factor from the simulation at $\varphi=0.50$
and $\varphi=0.58$, where we could get reasonable statistics for this
quantity. The MCT calculations then proceed by a linear interpolation between
these two cases to approximate $S(q)$ at nearby values of $\varphi$.
The critical nonergodicity parameters $f^c(q)$ thus obtained are shown in
Fig.~\ref{simsq-fc}. They
agree well for $qd\ge6$, lending confidence to the PY-based discussion
of the correlation functions. Smaller $q$ have been omitted from the figure
since there, differences can be seen that are related to insufficient
statistics for the simulated $S(q)$ at small wave numbers.
The results for the exponent parameter $\lambda$ also do not differ
significantly between the two calculations. We get $\lambda\approx0.735$
in the PY-based case \cite{Franosch1997}, and $0.727\le\lambda\le0.773$
based on the simulated $S(q)$, the latter value depending somewhat on the
discretization used.
The values for the critical packing fraction, however, differ between the two 
calculations: instead of $\varphi_{\text{MCT}}^c\approx0.516$, we get
$\varphi_{\text{MCT}}^c\approx0.585$ when using the simulated data to obtain $S(q)$.
Note that this makes this MCT result almost
coincide with what has been reported for
colloidal realizations of a hard-sphere system \cite{Megen1993}. Such
agreement is accidental, particularly because the value
$\varphi_{\text{sim}}^c$ extracted from our
simulations is even higher, but it demonstrates that the approximations used
for $S(q)$ need critical assessment.
Let us also note that the findings described here do not completely agree
with similar results reported in Ref.~\cite{Foffi2004}. There, the same
qualitative trend for $\varphi^c_{\text{MCT}}$ was found for a hard-sphere system,
and as well for two binary hard-sphere mixtures. But in this case, the
values for $f^c(q)$ based on the simulated structure-factor input
differed notably from those calculated within the PY
approximation, while we find no significant difference in this quantity.
In principle, our simulation-based results for $f^c(q)$ have no reason to be
closer to the PY results than the simulation-based results from
Ref.~\cite{Foffi2004}, since we use a slightly polydisperse soft-sphere
system, while in Ref.~\cite{Foffi2004}, strictly monodisperse hard spheres
have been simulated, at the cost of having to extrapolate to the desired
high densities.

The results shown in Fig.~\ref{simsq-fc} suggest that we may proceed in
the following discussion by basing all MCT results on the Percus-Yevick
approximation for $S(q)$. While this will make an adjustment of packing
fractions $\varphi_{\text{MCT}}$ necessary, it has the advantage of giving a
first-principles theory to compare the simulation data to. In particular,
the results presented above point out that neither the shape and strength
of the $\alpha$ relaxation, nor the asymptotic shape of the correlators in
the $\beta$-scaling regime will change much between the PY-based results and
those based on the simulated structure factor.

Before we embark on the comparison of the intermediate scattering functions
with the `full' MCT solutions, let us first analyze the simulation data
according to the asymptotic predictions of MCT, in oder to identify the time
window where MCT should certainly be applicable.

\subsection{Identification of Structural Dynamics}

\begin{figure}
\includegraphics[width=.75\linewidth]{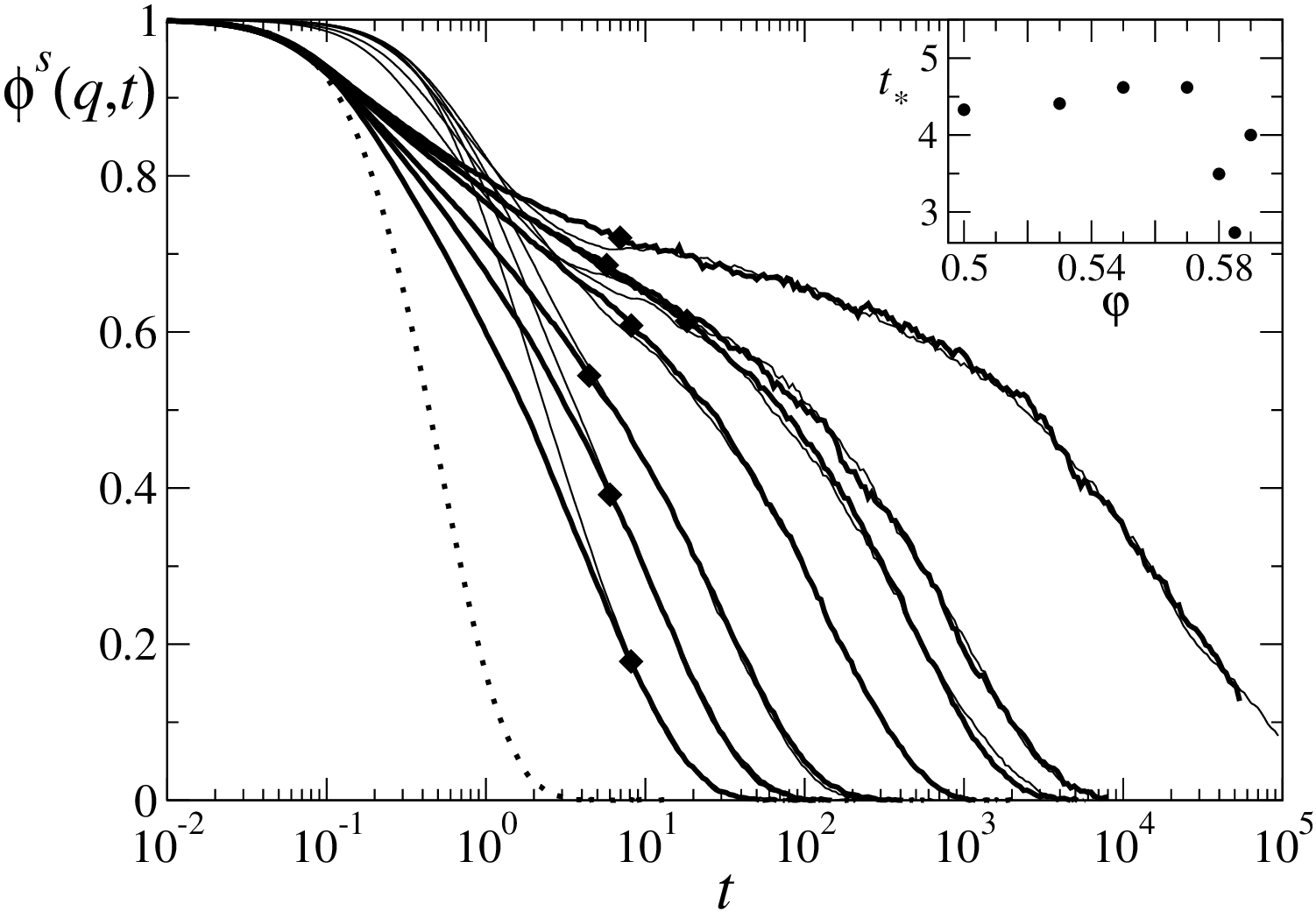}
\caption{\label{bd-nd-comp}
  Simulation results for $\varphi=0.50$, $0.53$, $0.55$, $0.57$, $0.58$,
  $0.585$, $0.59$ (from left to right) at wave vector $qd=7.8$. Heavy solid
  lines are the results using Brownian dynamics, thin lines the results for
  Newtonian dynamics. For the latter curves, times $t$ have been multiplied
  by factors $t_*$ given in the inset. The dotted line is the BD result
  for $\varphi=0.01$, indicating the dilute limit of the correlation function.
  The solid diamonds indicate the
  points where Newtonian and Brownian dynamics results start to agree at
  a $2\%$ level.
}
\end{figure}

In Fig.~\ref{bd-nd-comp}, results of the simulations are shown for
different packing fractions $\varphi$. A wave vector $qd=7.8$ close to the
first peak in the static structure factor has been chosen.
Different values of $q$ show qualitatively similar scenarios.
The thick solid lines in the figure are the simulation results for
`Brownian' dynamics simulations. Upon increasing $\varphi$, one observes
the emergence of a two-step relaxation process at times long compared with
typical liquid time scales. A typical relaxation curve for the dilute case,
is exemplified by the dotted line in the figure, showing the BD simulation
result for $\varphi=0.01$. From this, we read off a `microscopic' relaxation
time for the short-time relaxation of $t\approx1$.
The slow two-step relaxation pattern is usually
referred to as structural relaxation and is a precursor of the
approach to a glass transition at some $\varphi^c$. The scenario has been
observed repeatedly in similar systems. In our simulations, we are able to
follow the structural relaxation scenario for up to about five orders of
increase in the relaxation time.

MCT predicts that the structural relaxation becomes, up to a common time
scale $t_0$, independent on the
type of microscopic motion that governs the relaxation at short times.
To demonstrate that this is the case, Fig.~\ref{bd-nd-comp} shows as thin
lines the simulation results using Newtonian short-time dynamics. The data
have been scaled in $t$ in order to match the BD data at corresponding
packing fractions and at long times. Indeed, then the relaxation curves
match within our error bars at times $t>10$, indicated by the diamond symbols
in Fig.~\ref{bd-nd-comp}. Only at shorter times, the
regime of non-structural relaxation can be identified by the different
shapes of the BD and ND curves.
According to MCT, the scale factor $t_*=t_0^{\text{BD}}/t_0^{\text{ND}}$
used to match the BD and ND data at long times should be a smooth function
of $\varphi$, given by a constant in leading order close to $\varphi^c$.
For our simulation results,
the values are as shown in the inset of Fig.~\ref{bd-nd-comp}; they
are compatible with a constant shift $t_*\approx4.25$ within error bars.
Only at $\varphi=0.58$ and $\varphi=0.585$ do we note a stronger deviation,
the reason of which is unclear.
The overall variance in $t_*(\varphi)$ is comparable to the one found in a
similar study of a binary Lennard-Jones mixture \cite{Gleim1998}.

We conclude that the time window $t>10$ deals with structural relaxation
and thus comprises the regime where MCT predictions can be tested. At shorter
times, deviations from those predictions must be expected. There are
indications from theory \cite{Franosch1998,Fuchs1999b} and ND simulations
\cite{Kob2002} that these deviations are larger in ND.
We thus primarily discuss the BD data, which prove to be simpler to
understand within an MCT description.

\subsection{$\alpha$-process analysis}\label{alpha-analysis}

\begin{figure}
\includegraphics[width=.75\linewidth]{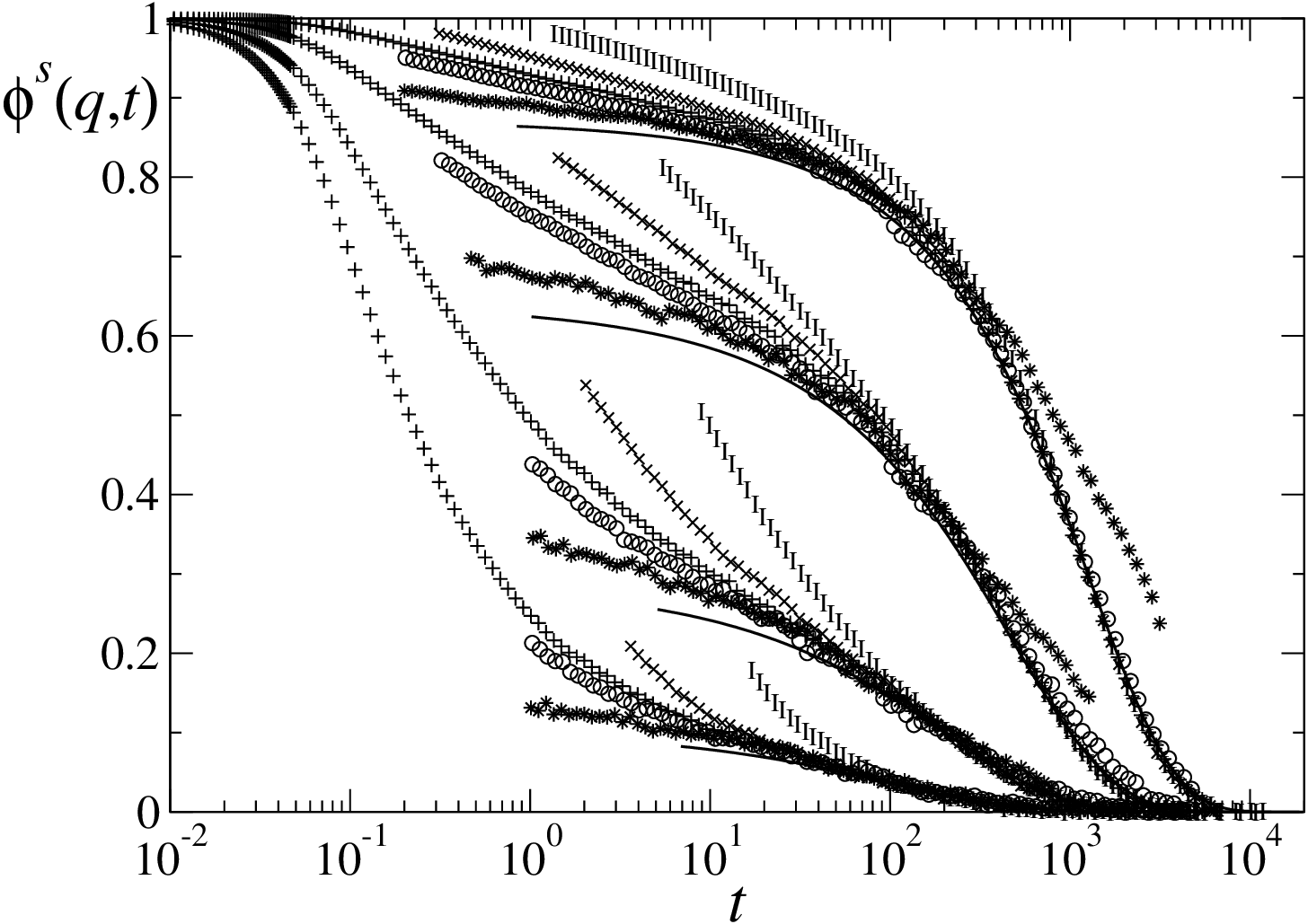}
\caption{\label{alphamaster}
  Comparison of the BD data for $\varphi=0.58$ (plus symbols) at wave vectors
  $qd=4.0$, $7.8$, $13.8$, and $19.8$ (from top to bottom).
  The long-time part of the data for $\varphi=0.55$ (I symbols),
  $0.57$ (crosses), $0.585$ (circles), and $0.59$ (star sybmols) is also
  shown, scaled in $t$ to match the long-time part of the $\varphi=0.58$ data.
  Solid lines are the MCT master curves for shifted $q_{\text{MCT}}$ (see text
  for details).
}
\end{figure}

The second step of the two-step relaxation process, i.e.\ the decay of
the correlators from their plateau value, is referred to as the
$\alpha$ process. A prediction of MCT is that the shape of this
$\alpha$ relaxation becomes independent on $\varphi$ in the limit
$\varphi\to\varphi^c-0$. Thus, scaling the correlation functions for
given $q$ and different $\varphi$ to agree at long times should
collapse the data onto a master curve.
Figure~\ref{alphamaster} demonstrates the validity of $\alpha$ scaling for the
BD data at several wave vectors between $q=4.0$ and $19.8$.
The scaling works as expected from the
MCT discussion of the HSS \cite{Franosch1997} for $\varphi\le0.58$.
The closer a state is to $\varphi^c$, the larger is the window where the
correlator follows the $\alpha$-master function. The increase of the
$\phi(q,t)$ above the master functions at shorter times is connected to the
$\beta$ process, discussed below. For $\varphi=0.59$, $\alpha$ scaling
breaks down at long times, $t\ge500$. The reason for this is unclear, and
cannot be understood within MCT. As observed by the orientational order
parameter $Q_6$, the system did not show appreciable trends to
crystallization in any of the analyzed simulation runs. Also for
$\varphi=0.585$, some
deviations from $\alpha$ scaling can be seen, particularly at $q=7.8$ and
at around $t=1000$, which are not in agreement with the pre-asymptotic
corrections to MCT $\alpha$ scaling. But in this case, the deviations are
less pronounced than those at $\varphi=0.59$.

\begin{figure}
\includegraphics[width=.75\linewidth]{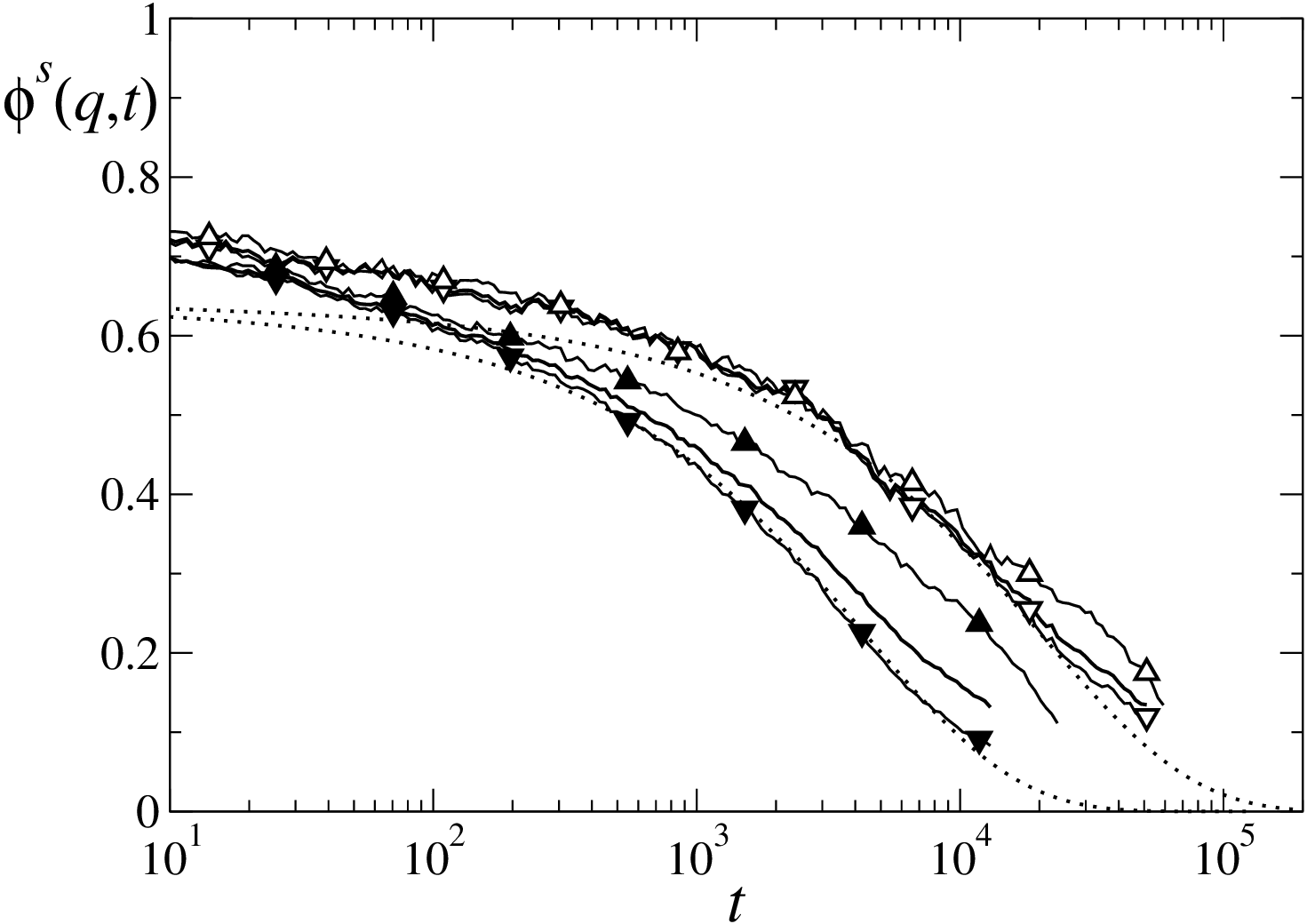}
\caption{\label{averages}
  Demonstration of the variability between different simulation runs for
  the ND and BD simulations at $\varphi=0.59$. Data is shown for $qd=7.8$,
  with open (filled) symbols denoting BD (ND) results. Triangles indicate
  averages over a small subset of the data ($8$ out of $30$ sets for BD;
  $25$ out of $70$ for ND) only; inverted triangles are the averages over the
  remaining data sets. The solid lines without symbols are the total
  averages. Dotted lines indicate the time-scaled MCT $\alpha$-master
  functions.
}
\end{figure}

The behavior of the long-time dynamics at these two densities,
$\varphi=0.585$ and $0.59$ is not fully understood. We have tried to
improve the statistical averaging by increasing the number of simulation
runs. However, there appear to be two subsets among the runs: one where the
$\alpha$-scaling violation is very pronounced, and one where the correlators
instead follow the scaling behavior much closer. This happens in both
the ND and the BD simulations, although the effect is more clearly seen
in the ND case. Out of the $30$ data sets we have averaged in the BD case
for $\varphi=0.59$, only $8$ show the scaling violation; in the ND case
we have averaged over $70$ sets, with $25$ of them deviating from scaling.
While Fig.~\ref{alphamaster} shows the data averaged over all simulation runs,
Fig.~\ref{averages} demonstrates the variation in $\alpha$-time scale
between the two types of data sets, obtained by restricting the averaging
to the number of data sets specified above. While we have no a priori
justification to modify the averaging procedure in any way, it allows us to
point out, that possibly some `rare' events take place in the system at
these high densities, which we cannot classify as crystallization events
on the basis of $Q_6$, but which modify the dynamical long-time behavior in
a distinct way. For the ND data, the $\alpha$-time scale varies
by a factor of $2.5$ between the two cases. In the BD data, the effect
is less pronounced, but still gives a factor of about $1.6$.
As the dotted lines in Fig.~\ref{averages} demonstrate, the majority of the
data sets follows the predicted scaling rather closely, whereas a the
remaining ones show significantly slower decay.

We have tried to analyze this finding further by looking at the distribution
of squared displacements exhibited by all the particles,
$P_{\text{MSD},t_*}(\delta r^2)$, and its correlation with particle size.
Here, $t_*$ is a fixed time, and the distribution is defined such that
$\int P_{\text{MSD},t_*}(\delta r^2)dr^2=\delta r^2(t_*)$. We have fixed
$t_*$ such that $\delta r^2(t_*)=1.25d^2$. The distribution $P_{\text{MSD}}$
develops a non-Gaussian peak centered
around its average value, whose width increases upon increasing the
packing fraction. In some cases, we did observe a two-peaked distribution
at $\varphi=0.59$, signifying that a certain amount of particles is displaced
significantly less than average, i.e.\ that populations of `fast' and `slow'
particles develop. This might be connected to the `rare events'
mentioned above, but we point out that this finding is unstable against
improving the average over an increased number of simulation runs.

From the $\alpha$-scaling plot, Fig.~\ref{alphamaster},
we infer the regime of $\alpha$-relaxation
dynamics. Note that for $\varphi=0.58$, deviations from the $\alpha$-master
curve due to $\beta$ relaxation are seen almost up to $t\approx100$. Those
will be analyzed later.
Also shown in Fig.~\ref{alphamaster} are the MCT $\alpha$-master functions.
If evaluated at the same $q$ as the simulation data, the description of
the long-time dynamics is unsatisfactory, because the calculated stretching
of the relaxation is too small. If we account for this error by shifting
$q_{\text{MCT}}$ used in the calculations to higher values, we get good agreement,
cf.\ the solid lines in Fig.~\ref{alphamaster}.
Note that the deviations from the $\alpha$-master curve set in at a time
later than that where the ND and BD simulation results begin to overlap:
e.g., the $q=7.8$ curve follows the $\alpha$-master curve only for times
$t\gtrsim100$, as can be inferred from Fig.~\ref{alphamaster}. Still, the
BD and ND curves for that state collapse within our error bars already for
$t\gtrsim20$, cf.\ Fig.~\ref{bd-nd-comp}. This underlines that the regime of
structural relaxation identified in Fig.~\ref{bd-nd-comp} is larger than
that of the $\alpha$-decay regime observed in Fig.~\ref{alphamaster}, i.e.\
that both simulation data sets show some extent of the MCT $\beta$-relaxation
window.

The values of $q_{\text{MCT}}$ used in the fits of Fig.~\ref{alphamaster}
are $q_{\text{MCT}}=5.0$ ($9.13$, $10.3$, $15.13$, $18.3$, $20.87$) for $q=4.0$
($7.8$, $9.0$, $13.8$, $17.0$, $19.8$). These fit values are suggested by the
analysis of the full curves pursued below, cf.\ Sec.~\ref{mct-analysis}.
Comparing the fitted wave-vector values $q_{\text{MCT}}$ to those of the simulations,
deviations in $q$ are in the range $10\%$ to $17\%$, except for $q=4.0$,
where a $25\%$ deviation is needed to describe the $\alpha$-master function.
The way we have adjusted $q_{\text{MCT}}$ ensures that the stretching of the
correlators is described correctly. In contrast, a fit of the plateau values
with the $\alpha$-master functions is difficult, since the latter are
still not clearly visible in the simulation data even at $\varphi=0.59$.
This will become more apparent in Sec.~\ref{mct-analysis}.

In all cases, the fitted wave-vector values are larger than the actual values,
$q_{\text{MCT}}\ge q$. Since the $f^{s,c}(q)$ giving the plateau values
decrease monotonically from unity at $q=0$ to zero at $q\to\infty$, this
fit result is equivalent to stating that the MCT-calculated plateau values
appear too high. Furthermore, the half-width of the $f^{s,c}(q)$ distribution
is an estimate for the inverse localization length of a tagged particle.
Hence the fit suggests that MCT underestimates the localization length
of a tagged particle in the system slightly.
There are two obvious reasons for such a mismatch in length scales:
first, the softness of the particles in the simulation might, especially at
high densities, give rise to some amount of particle overlap not possible in
the HSS, rendering the effective localization of the particles slightly
larger. According to Heyes \cite{HeyesJCP}, the soft-sphere system used
in our simulations can be well described within the hard-sphere limit and an
effective hard-sphere diameter $d_{\text{eff}}=
\int_0^\infty(1-\exp[-\beta V_c(r)])dr\approx d(1+\gamma_e/36)\approx1.016$
(where $\gamma_e\approx0.577$ is Euler's constant), which differs from $d=1$
by less than $2\%$.
But one has to keep in mind that the convergence of
increasingly steep soft-sphere potentials towards the hard-sphere limit
can be quite slow for the transport properties of the system
\cite{HeyesPowles}.
Second, a difference in packing fractions between the simulation and the
MCT calculation might become important in this respect. This
arises, because within the PY approximation for the DCF, the MCT master
curve is evaluated at the corresponding value for the critical packing
fraction, $\varphi^c\approx0.516<\varphi^c_{\text{sim}}$.
As was pointed out in connection with Fig.~\ref{sq-comp}, such a
mismatch in $\varphi$ will affect the average particle distances, and thus
an overall length scale. But since $\varphi^c_{\text{MCT}}<\varphi^c$ one would expect
this to lead to an overestimation of the critical localization length,
contrary to what we observe.

\begin{figure}
\includegraphics[width=.75\linewidth]{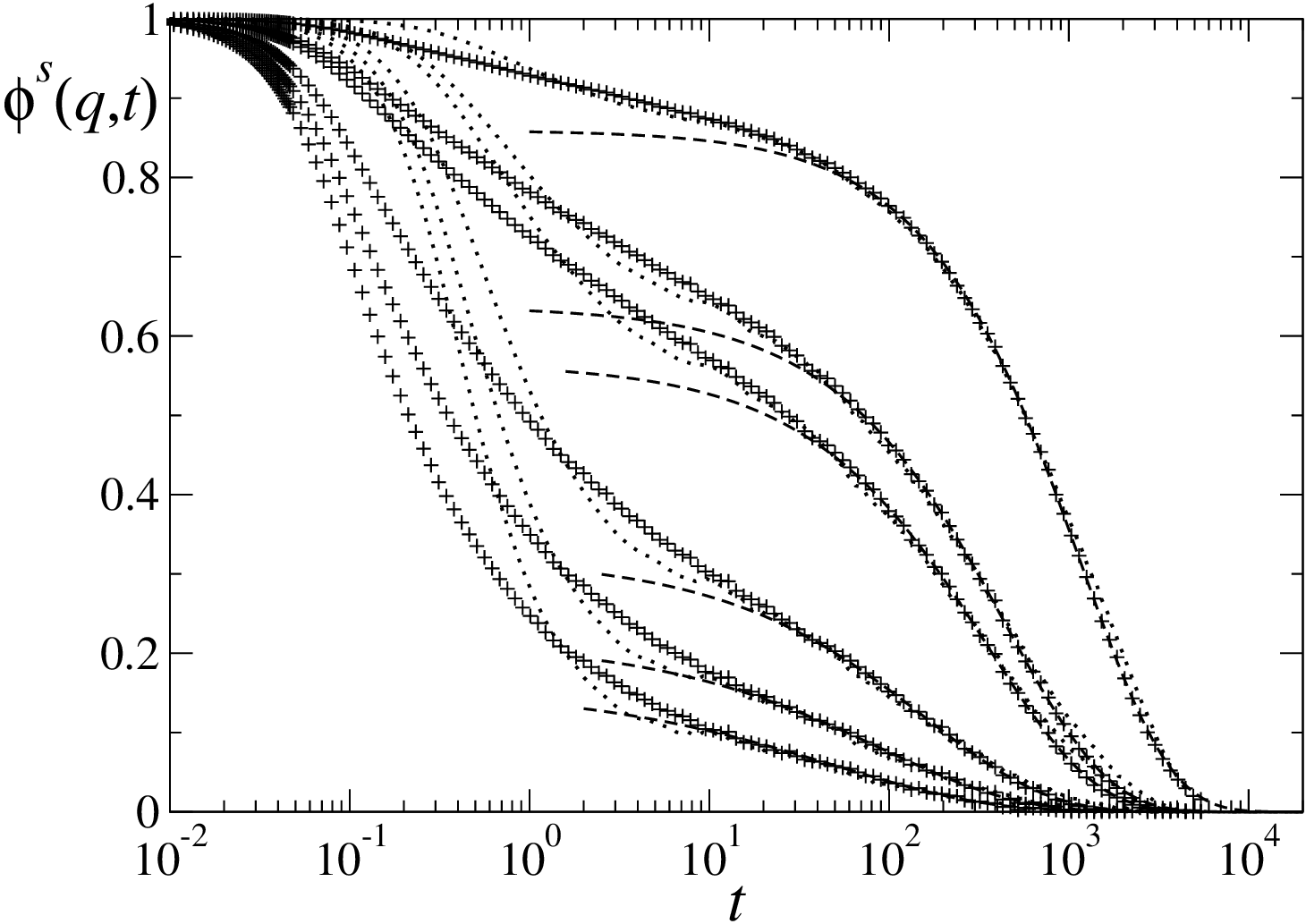}
\caption{\label{kwwdemo}
  Example for Kohlrausch fits to the simulation data at $\varphi=0.58$
  (symbols: Brownian dynamics, dots: Newtonian dynamics),
  $qd=4.0$, $7.8$, $9.0$, $13.8$, $17.0$, and $19.8$.
  The fit range was $t\ge55$.
}
\end{figure}

Traditionally, stretched-exponential (Kohlrausch) laws,
\begin{equation}\label{kww}
  \phi^s(q,t)\approx A(q)\exp[-(t/\tau(q))^{\beta(q)}]\,,
\end{equation}
are known to give a good empirical description of the $\alpha$ relaxation.
Here, $A(q)$ is an amplitude factor, $\tau(q)$ the Kohlrausch time-scale
of the $\alpha$ relaxation, and $\beta(q)<1$ is called the stretching
exponent. These parameters in general depend on the observable under study,
and in particular on the wave vector $q$.
Fig.~\ref{kwwdemo} demonstrates that the Kohlrausch laws can also be used to
fit the $\alpha$-relaxation part of our simulation data. The figure shows
as an example the state $\varphi=0.58$ for various wave vectors.
One problem of the stretched-exponential analysis of the data is that the
three parameters of the fit have a systematic dependence on the
fit range. In particular, one has to restrict the fitting to such large $t$
that only $\alpha$ relaxation is fitted. For the fits shown, this range was
chosen to be $t\ge55$. The data
deviate from the fitted Kohlrausch functions significantly only at shorter
times; but there is a trend that these deviations set in just about the
boundary of the fit range. This still holds if the fit is restricted to
larger $t$ only, and judging from the fit quality for the remaining relaxation
alone, one cannot determine the optimal choice of the fit range. It is thus
particularly difficult to extract the regime of $\alpha$ relaxation from
the Kohlrausch fits alone. On the other hand, from the MCT fits shown in
Fig.~\ref{alphamaster} we expect corrections due to $\beta$ relaxation to set
in at about $t\approx100$. This in principle gives an indication of the maximum
fit range to chose. Yet, Kohlrausch fits restricted to $t\ge100$ did lead to
an unsatisfactory scatter in the fit parameters $A(q)$ and $\beta(q)$. Thus
an analysis of the $\alpha$ relaxation using Kohlrausch fits will
erroneously include parts of the $\beta$ relaxation.

\begin{figure}
\includegraphics[width=.75\linewidth]{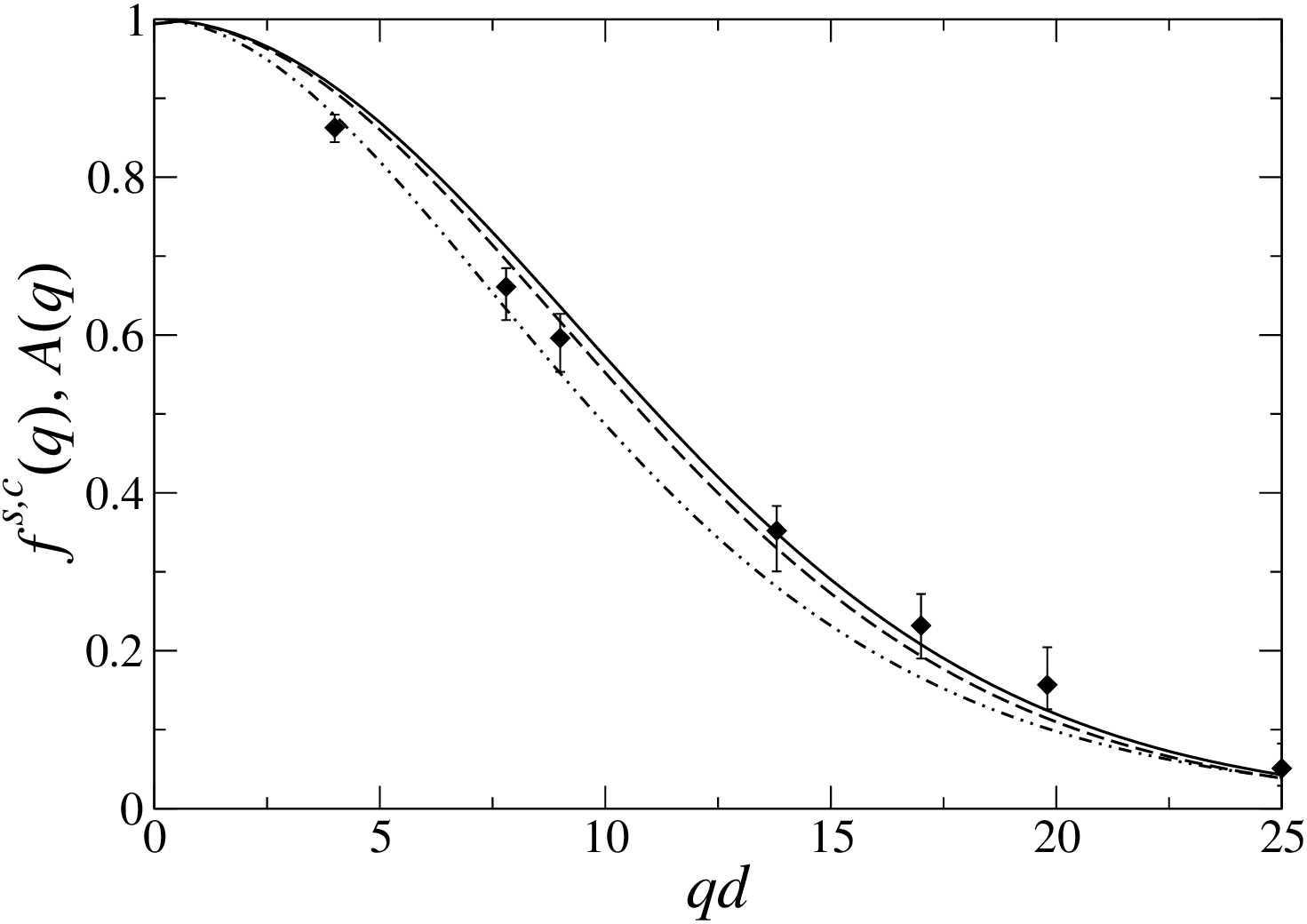}
\caption{\label{fcomp}
  Critical nonergodicity parameter $f^{s,c}(q)$ calculated from MCT for the
  one-component hard-sphere system with PY approximation (solid line),
  and for a five-component polydisperse system (dashed line). Symbols are
  the amplitudes $A(q)$ from Kohlrausch fits to the data, where error bars
  estimate deviations depending on $\varphi$ and BD/ND.
  The dash-dotted line indicates $f^{s,c}(q)$, but transformed with the
  wave-number shift applied in the discussion of the dynamical data
  (see text for details).
}
\end{figure}

This trend can be also identified comparing the obtained Kohlrausch
amplitudes $A(q)$ with the plateau values $f^{s,c}(q)$. This comparison is
shown in Fig.~\ref{fcomp}, where the MCT results for $f^{s,c}(q)$ are
included. They have again been determined using the PY approximation
to $S(q)$, but in agreement with Fig.~\ref{simsq-fc}, the values for
$f^{s,c}(q)$
determined from MCT calculations based on the simulated $S(q)$ are
indistinguishable from the ones shown on the scale of Fig.~\ref{fcomp}.
For Kohlrausch fits to the $\alpha$-master function, and
more generally to the $\alpha$-relaxation regime of the correlators only,
$A(q)\le f^{s,c}(q)$ should hold. Recalling the wave-number adjustment used
in Fig.~\ref{alphamaster}, we should even have $A(q)\le f^{s,c}(q_{\text{MCT}}(q))$.
This latter curve is included in Fig.~\ref{fcomp} as the dash-dotted line,
where the mapping $q\mapsto q_{\text{MCT}}$ was extended from the set of $q$
analyzed in this text to all $q$ via a quadratic interpolation.
The relation $A(q)\le f^{s,c}(q)$ is clearly violated for the fits here,
especially at high $q$. It shows that the distinction between the $\alpha$
and $\beta$ regimes from the simulation data is difficult; increasingly so
with increasing wave number.

It is reassuring that the Kohlrausch fits to BD and ND data yield values
quite close to each other, apart from an overall shift in $\tau(q)$. This
holds, as long as the fit ranges are chosen such as to fit approximately
the same part of the relaxation. In Fig.~\ref{kwwdemo},
the ND curves have been added, again shifted
by a scaling factor in $t$ given in the inset of Fig.~\ref{bd-nd-comp}.
Note that, while in the ND curves one can identify a plateau from
the data better than from the BD ones, still the KWW fits have a trend
to give too high values of $A(q)$. Thus one has to be careful when
extracting plateau values from the simulation data by such an analysis,
even if the data seem to give a clear indication of the plateau.

\begin{figure}
\includegraphics[width=.75\linewidth]{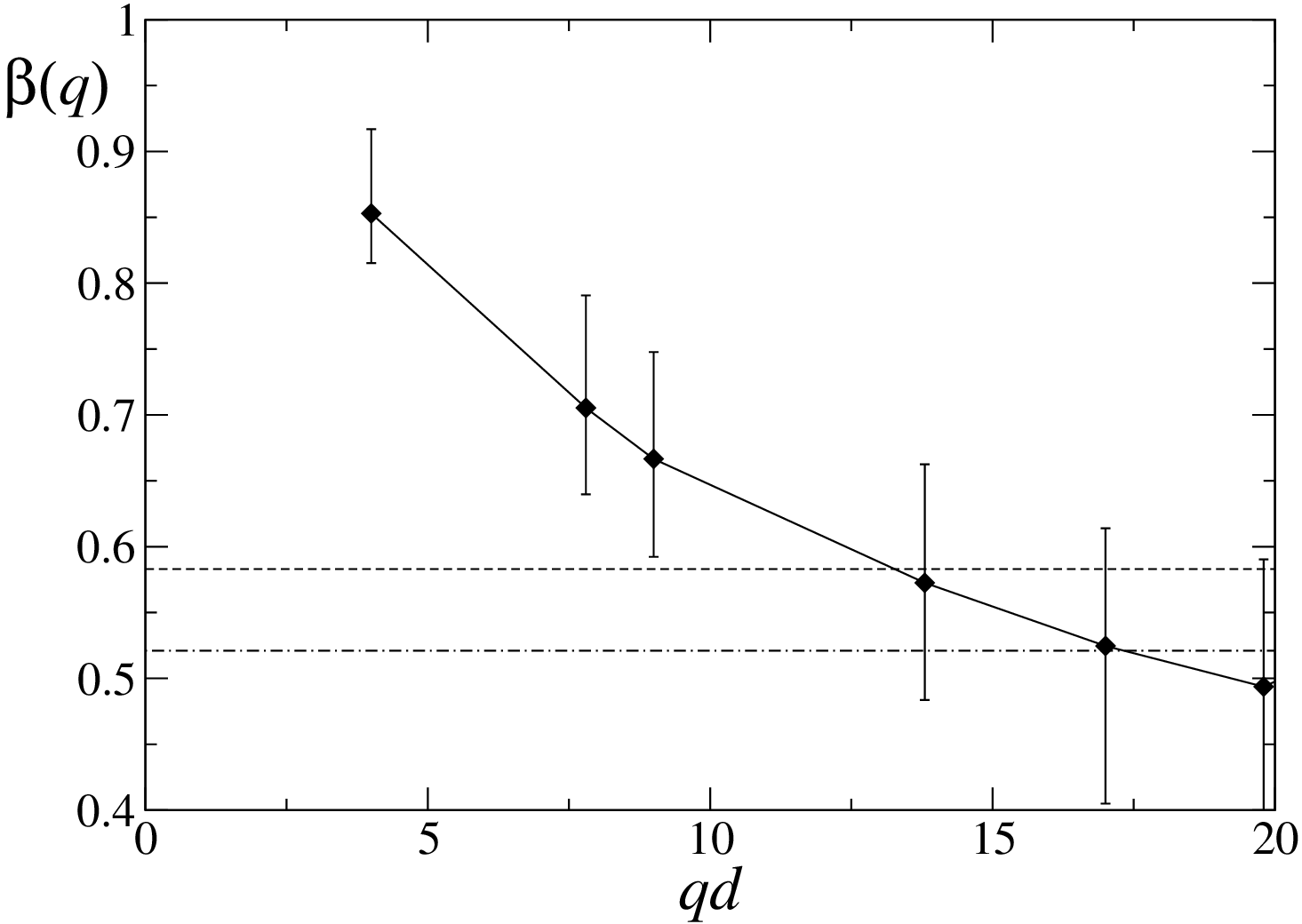}
\caption{\label{kwwbeta}
  Stretching exponents $\beta(q)$ from fits to the BD simulation data
  at $\varphi=0.58$ using Kohlrausch laws, Eq.~\protect\eqref{kww}.
  Error bars indicate deviations estimated from
  fits to ND data and to $\varphi=0.57$. The dashed horizontal line indicates
  the $b$ value calculated from MCT using the Percus-Yevick approximation
  for $S(q)$, $b\approx0.583$, and the dash-dotted line is $b$ as determined
  from MCT with simulation-data input for $S(q)$, $b\approx0.521$.
}
\end{figure}

The stretching exponents $\beta(q)$ from the KWW fits are shown in
Fig.~\ref{kwwbeta}. Again, we have included error bars indicating the
deviations arising from fits to different $\varphi$ or to ND as opposed to
BD simulation data. $\beta(q)$ increases with decreasing $q$, and this
increase is compatible with $\beta(q)\to1$ for $q\to0$, as expected from
theory \cite{Fuchs1992b}. According to MCT, $\beta(q)$ should approach the
von~Schweidler exponent $b$ as $q\to\infty$ \cite{Fuchs1994}. The value of
$b$ is calculated from the MCT exponent parameter $\lambda$, and for the
HSS using the PY-DCF is $b\approx0.583$,
shown as a dashed line in Fig.~\ref{kwwbeta}. We observe
that the fitted $\beta(q)$ fall below this value for large $q$, even if the
fits are less reliable there, due to the low amplitudes $A(q)$ at high $q$.
To estimate the error of the theory prediction for $b$, we have also calculated
this exponent from MCT using the simulated data as input for $S(q)$.
According to the values of $\lambda$ reported in connection with
Fig.~\ref{simsq-fc}, we get
$b\approx0.56\pm0.04$. The lower bound for $b$ thus obtained
is indicated in Fig.~\ref{kwwbeta} as the dash-dotted line. Taking into
account this uncertainty, the behavior of the fitted $\beta(q)$ agrees well
with what is expected from theory. In the further discussion, we will fix
$\lambda$ to its value derived from the PY approximation, $\lambda=0.735$.
Since the shape of the correlation functions in the $\beta$ regime is in the
asymptotic limit fixed by $\lambda$, some deviations in the fits described
below are to be expected in this time window.

\subsection{Analysis of $\alpha$-relaxation times}

\begin{figure}
\includegraphics[width=.75\linewidth]{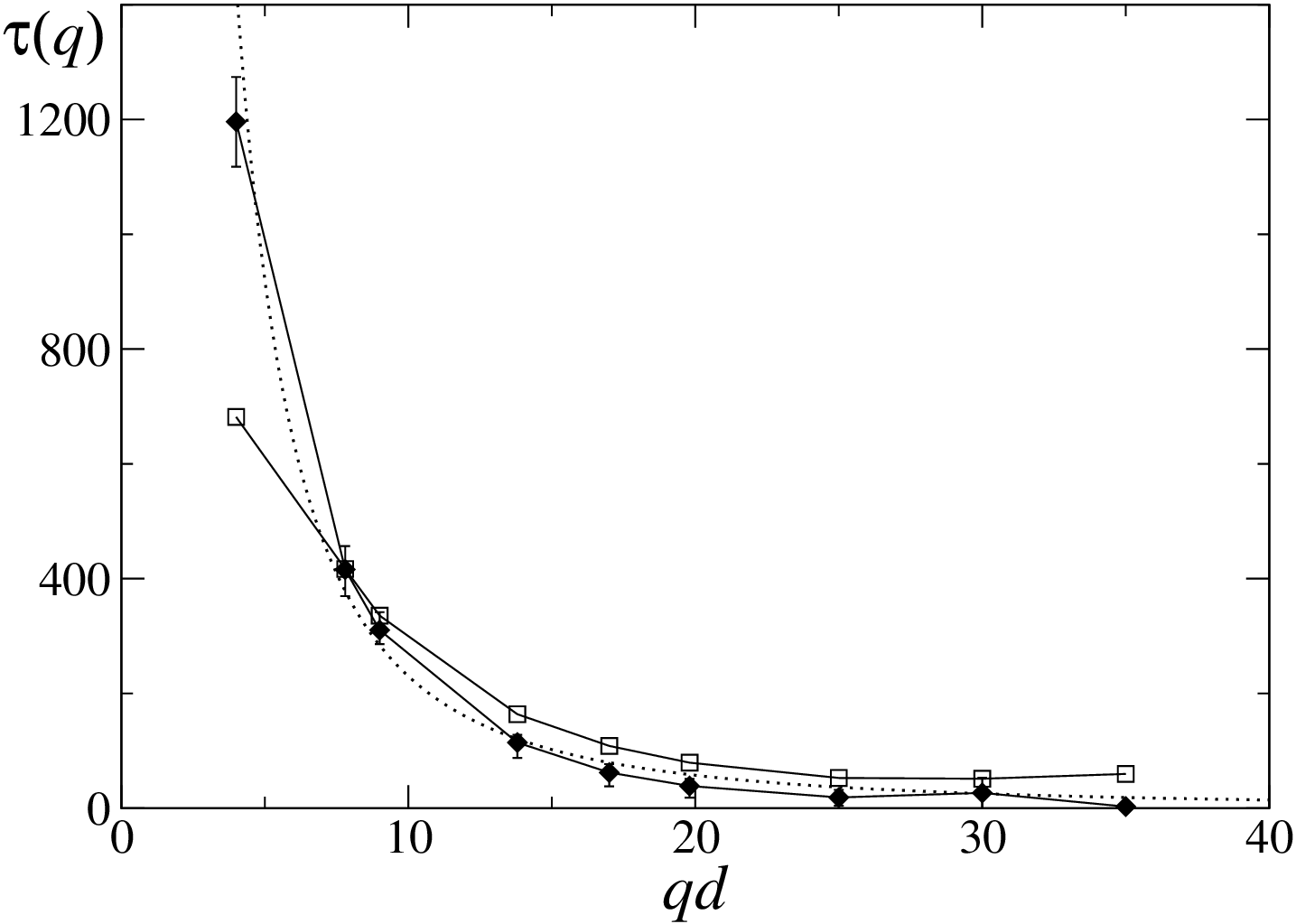}
\caption{\label{tauq}
  $\tau(q)$ from Kohlrausch fits, Eq.~\protect\eqref{kww}, to the BD data at
  $\varphi=0.58$ (diamonds) and at $\varphi=0.59$ (squares; scaled by a
  factor of $0.021$). Error bars estimate the uncertainty from the fits,
  see text for details. The dotted line shows a $1/q^2$ law.
}
\end{figure}

The $q$-dependence of the $\alpha$-relaxation times at fixed $\varphi$ can
best be analyzed from the $\tau(q)$ extracted from Kohlrausch fits. In
Fig.~\ref{tauq}, we report values for $\tau(q)$ for such fits to the BD data at
$\varphi=0.58$ as the diamond symbols. If one instead fits the ND data, or
data at $\varphi=0.585$ or $0.57$, the $q$-dependence is the same up to a
prefactor and up to small deviations. These deviations are indicated by the
size of the error bars in Fig.~\ref{tauq}. The data closely follow a $1/q^2$
dependence for small $q$, indicated by the dotted line.
This is in agreement with earlier MCT predictions for the hard-sphere
system \cite{Fuchs1992b}. For $q\to\infty$, one expects from MCT $\tau(q)\sim
q^{-1/b}$. But since $b$ is close to $1/2$, we cannot distinguish this
behavior reliably from a $1/q^2$ law due to the noise of the data at
large $q$.

Fits to the BD data at $\varphi=0.59$ reveal significant deviations from
the behavior at $\varphi\le0.585$ at small $q$. This can be deduced from the
square symbols in Fig.~\ref{tauq}. They have been scaled by a constant factor
in order to match the value obtained from the $\varphi=0.58$ fit at
$q=7.8$, since there the MCT analysis works best, as will be shown below.
At smaller $q$, the increase of $\tau(q)$ with decreasing $q$ is suppressed
for the $\varphi=0.59$ data in comparison to the variation in $\tau(q)$
observed for smaller $\varphi$.

\begin{figure}
\includegraphics[width=.75\linewidth]{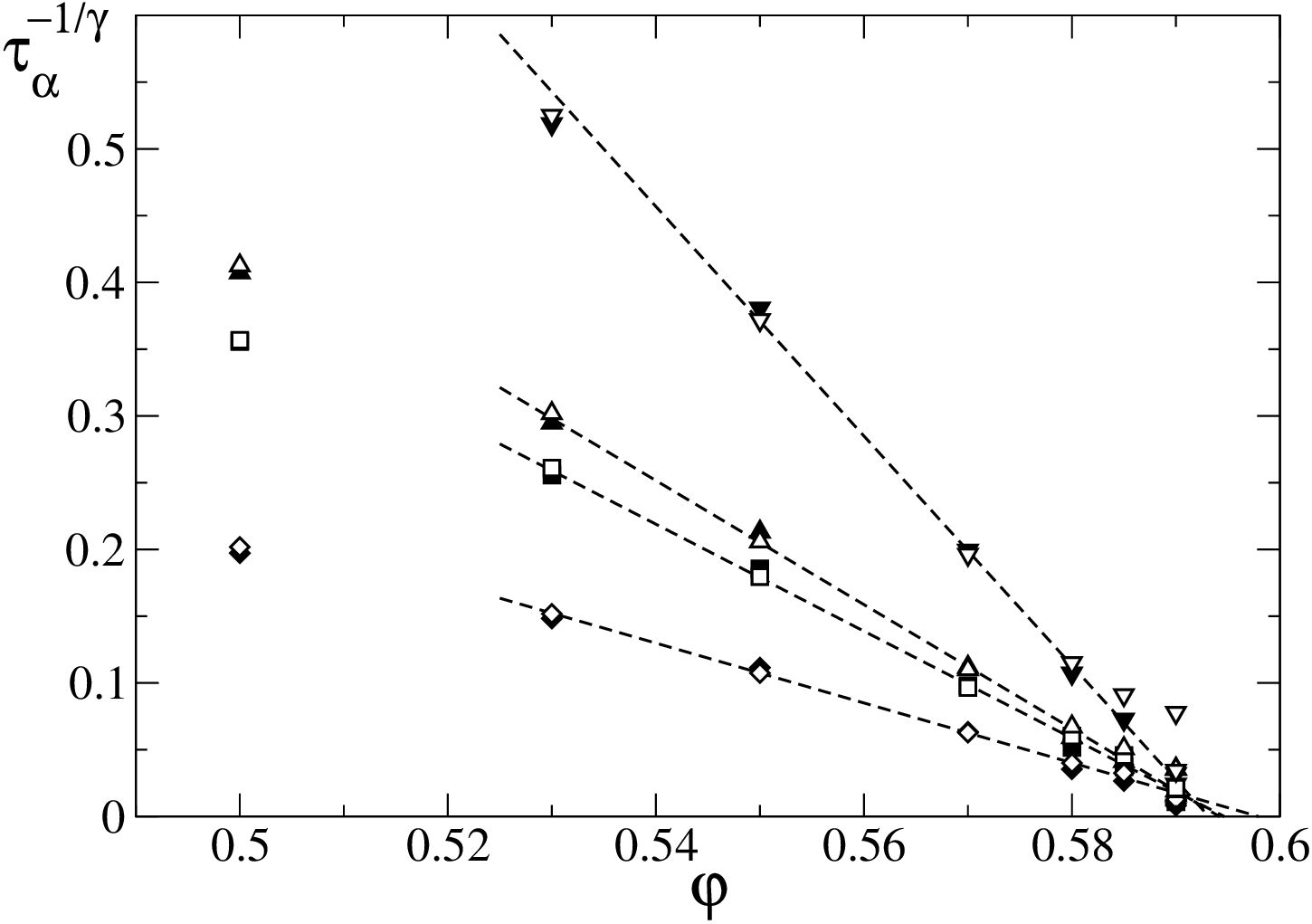}
\caption{\label{taualpha}
  Plots of $\tau_\alpha^{-1/\gamma}$ for wave vectors $qd=4.0$ (diamonds),
  $7.8$ (squares), $9.0$ (up triangles), and $13.8$ (down triangles); using
  $\gamma=2.46$.
  (a) Results for the simulation data using Brownian dynamics (open symbols)
  and Newtonian dynamics (filled symbols). The values of $\tau$ for the latter
  have been multiplied by $4.5$ for this plot. Dashed lines are linear
  fits to the Brownian dynamics data in the range $0.54\le\varphi
  \le0.58$.
}
\end{figure}

For a discussion of the density dependence of the $\alpha$-relaxation
time, the $\tau(q)$ from the stretched-exponential fits are less reliable,
since the Kohlrausch fits suffer from larger uncertainties at lower $\varphi$,
where the $\alpha$- and $\beta$-regimes are even less well separated.
But we can operationally define a time scale $\tau_\alpha$ for the decay of
the correlation functions as the point where the correlators have decayed to
$10\%$ of their initial value, $\phi^s(q,\tau_\alpha)=0.1$. For small enough
$q$ where $f^{s,c}(q)$ is still much larger than $0.1$, $\tau_\alpha$ is a
useful measure for the $\alpha$-process time scale. In the asymptotic regime,
where $\alpha$ scaling holds, it follows the $\alpha$-scaling time defined
within MCT up to a constant. Thus MCT predicts a power-law divergence of
$\tau_\alpha$ close to $\varphi^c$ of the form $|\varphi-\varphi^c|^{-\gamma}$
for not too large $q$, $q\le15$, say. To test this prediction,
we plot in Fig.~\ref{taualpha} the quantity $\tau_\alpha^{-1/\gamma}$,
which should yield a straight line crossing zero at $\varphi^c$. Since the
region of validity of this asymptotic result is not known a priori, a
determination of the correct value of the exponent $\gamma$ on the basis of
such rectification plot suffers from large errors. Therefore, let us fix
$\gamma\approx2.46$, the value calculated by MCT using the PY approximation.
Due to the uncertainty in determining $\lambda$ mentioned above, slightly
different values of $\gamma$ cannot be ruled out. One gets $\gamma\approx2.66$
as an upper bound when using the upper bound for $\lambda$ given above for
the MCT result based on the simulated $S(q)$. This value of $\gamma$ is also
quite close to what one gets \cite{Fuchs1992b} using the Verlet-Weis-corrected
PY structure factor \cite{Verlet1972}.
Figure~\ref{taualpha} shows rectification plots for both
Brownian and Newtonian dynamics simulation data for $4.0\le q\le13.8$. For
the latter, the values of $\tau_\alpha$ have been multiplied by $4.5$,
consistent with the shift in the overall time scale, cf.\ inset of
Fig.~\ref{bd-nd-comp}. Fits to straight lines give $\varphi^c$ values that are
consistent with each other for $q\ge7.8$ and both microscopic dynamics, if
one restricts the fit range to high enough $\varphi$ and omits the highest
densities, where alpha scaling breaks down, $0.54<\varphi<0.58$ in our case.
From this, one gets $\varphi^c\approx0.594\pm0.001$. The data for
$q=4.0$ give a somewhat higher value, $\varphi^c\approx0.598\pm0.001$,
again the same for Newtonian and Brownian dynamics.
This difference is outside the error bars of the analysis and thus not
in accord with MCT. Since the discrepancy is the same for both types of
short-time dynamics, we conclude that indeed the structural relaxation
deviates from the MCT prediction systematically at small $q$.

The range of distances $\varepsilon$ to the critical point, in which we can
fit the time scales consistently with a power law, is roughly
$|\varphi-\varphi^c|/\varphi^c\le0.07$. This agrees with what is
expected from a discussion of the asymptotic MCT results for the hard-sphere
system \cite{Franosch1997}: For the time-scales extracted from the numerical
MCT results, we have to restrict the linear fit to $\varphi_{\text{MCT}}\ge0.48$,
where the critical point is $\varphi^c_{\text{MCT}}\approx0.516$. Below
$\varphi_{\text{MCT}}=0.48$, one finds deviations from the straight lines in the
rectification plot; typically the results fall below the asymptotic straight
line in such a plot. If one tries to fit a larger range in $\varphi_{\text{MCT}}$,
the thus estimated $\varphi^c$ will be higher than the correct one. For
example, we get $\varphi^c\approx0.519\pm0.0015$ when fitting in the range
$\varphi_{\text{MCT}}\ge0.4$. It is reassuring that the deviations from the linear
behavior seen in Fig.~\ref{taualpha} for the simulation results occur
in the same direction as found for the MCT results.

At large $q$ and at the highest packing fractions studied, the $\tau_\alpha$
from the simulations are systematically smaller than what is expected from the
power-law extrapolation. This suggests that the local relaxation dynamics of
the system very close to the transition would be faster than expected within
the theory. However, the full theory analysis presented below suggests the
opposite, indicating that at these high $q$, the operational definition of
$\tau_\alpha$ we have chosen for simplicity no longer works.

\subsection{$\beta$-process analysis}\label{betarelaxation}

We now test some of the predictions MCT makes for the $\beta$-relaxation
regime, where the correlators remain close to their plateau values. The
time window where the asymptotic solution holds, extends in $t$ upon control
parameters approaching the transition point, i.e.\ $\varphi\to\varphi^c$.
The leading deviation from the plateau value is of order $\sqrt{|\sigma|}$,
where $\sigma$ is called the distance parameter, and
\begin{align}\label{sigmaequation}
\sigma&=C\cdot\varepsilon\,, &
\varepsilon&=(\varphi-\varphi^c)/\varphi^c\,,
\end{align}
in leading order is the linearized distance in control-parameter space.
The leading-order asymptotic result is called factorization theorem, and it
can be written for the tagged-particle correlator as
\begin{equation} \phi^s(q,t)=f^{s,c}(q)+h^s(q)G(t)\,.\end{equation}
Here, $G(t)$ is a universal function depending only on the parameters
$\lambda$, $\sigma$, and a fixed `microscopic' time scale $t_0$. The
expansion is valid on a time scale $t_\sigma=t_0|\sigma|^{-1/(2a)}$ that
diverges upon approaching the transition point. On this time scale, all
wave-vector dependence is factorized off from the time dependence,
and contained in the critical amplitudes $h^s(q)$ and the plateau values
$f^{s,c}(q)$. All parameters can be calculated within MCT, but as we have
seen above, extracting
them from the simulation data is not straightforward. Kob et~al.\
\cite{Gleim2000} have introduced a test of the factorization property that
can be performed without fitting any of the $q$-dependent quantities to the
data: they considered
\begin{equation}\label{xeq}
  X(q,t)=\frac{\phi^s(q,t)-\phi^s(q,t')}
  {\phi^s(q,t')-\phi^s(q,t'')}\,. \end{equation}
If fixed times $t'$ and $t''$ are chosen inside the $\beta$ regime,
the factorization theorem gives $X(q,t)\equiv X(t)=x_1G(t)-x_2$ for $t$ inside
the $\beta$ regime, with constants $x_1$ and $x_2$ not depending on $q$.
Therefore, if the factorization theorem holds, plotting $X(q,t)$ for various
$q$ collapses the curves in the $\beta$ window, without the need for fitting
wave-vector dependent amplitudes and plateau values.

\begin{figure}
\includegraphics[width=.75\linewidth]{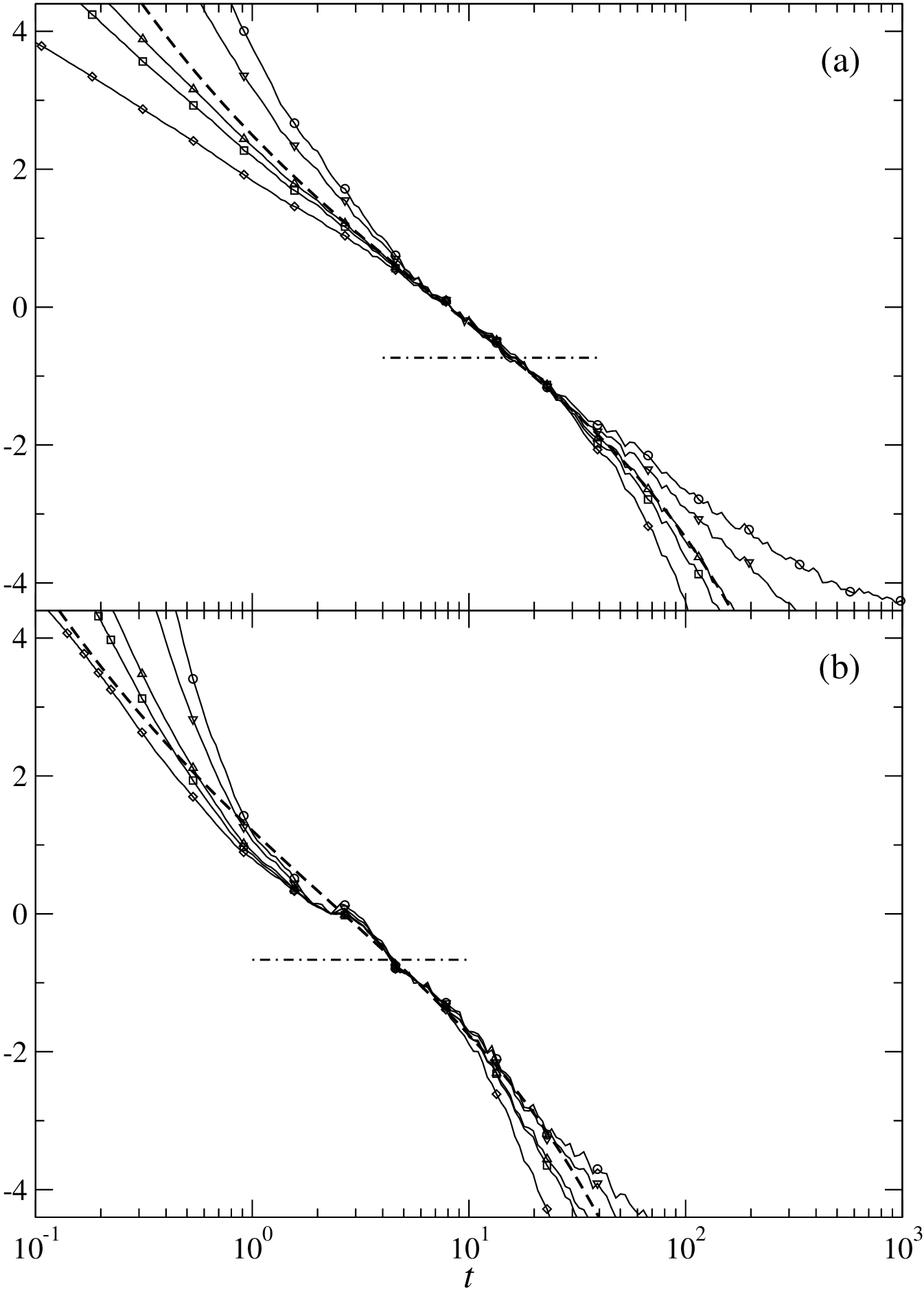}
\caption{\label{beta}
  (a) $\beta$ analysis of the BD simulation data at $\varphi=0.58$:
  the curves marked by symbols show $X(q,t)=(\phi^s(q,t)-\phi^s(q,t'))
  /(\phi^s(q,t')-\phi^s(q,t''))$ with $t'=8.234$, $t''=20.8075$. Wave vectors
  are $qd=4.0$ (diamonds), $qd=7.8$ (squares), $qd=9.0$ (up triangles),
  $qd=13.8$ (down triangles), and $qd=17.0$ (circles). The dashed line
  is the equivalent of the MCT $\beta$ correlator, see text for details.
  The dash-dotted line indicates the plateau value estimated from the
  root of the $\beta$ correlator.
  (b) Same for the ND simulation data, $t'=2.31$ and $t''=5.845$.
}
\end{figure}

We have performed this test for our simulation data for both BD and ND.
Figure~\ref{beta}a shows the results for the BD simulation at $\varphi=0.58$,
with $t'=8.234$ and $t''=20.8075$. One observes that the data nicely collapse
for $5\le t\le 40$. Additionally, the figure shows the $X(t)$ constructed from
the MCT $\beta$ correlator as a dashed line. Here, two constants $x_1$ and
$x_2$, and the time scale $t_\sigma$ have been fitted. The value of $\lambda$
has been taken from the theory as explained above, $\lambda=0.735$.
The same analysis is carried out for the ND data in Fig.~\ref{beta}b.
Here, we have fixed $t'_{\text{ND}}=2.31\approx t'/3.5$ and
$t''_{\text{ND}}=5.845\approx t''/3.5$, since at $\varphi=0.58$ the shift in
time scales between BD and ND is a factor of about $3.5$, cf.\ inset of
Fig.~\ref{bd-nd-comp}. Again the data collapse
in an intermediate window $2\le t\le 15$. The upper end of this window is
consistent with the one found for the BD analysis, i.e.\ $15\approx40/3.5$.
The lower end of the window where $\beta$ scaling holds for the ND data
is higher than what would correspond to the BD $\beta$ window.
Thus preasymptotic corrections are stronger in the ND case.
The fit using the MCT $\beta$ correlator is not as good as it is for the BD
case. Since the distance to the critical point does not change between BD and
ND, we have used the $t_\sigma$ determined from the above fit to the BD data
also here. Furthermore, since we have chosen $t'_{\text{ND}}$ and
$t''_{\text{ND}}$ in accordance with the values of $t'$ and $t''$ for the BD
analysis, the constants $x_1$ and $x_2$ should be the same; this is roughly
fulfilled by our fit.

The fits to both data sets have been performed
such as to obey the `ordering rule' for the corrections to $\beta$ scaling
\cite{Franosch1997}: a curve that falls below another one for times smaller
than the $\beta$ window, will also do so for time larger than the $\beta$
regime, since the corrections both at small and at large times are determined
by the same $q$-dependent correction amplitudes. Thus the ordering of wave
vectors on both sides of the scaling regime is preserved. This prediction of
MCT is fulfilled by the BD simulation data, as can be seen in Fig.~\ref{beta}a.
For the ND data, we cannot fulfill this ordering at both short and long times
with reasonable fit parameters. At shorter times, one finds that e.g.\
the $q=7.8$ and $q=9.0$ curves rise above the $\beta$-correlator curve,
in violation of the ordering rule. We thus conclude that at this point,
microscopic deviations set in for the ND simulations
that are stronger than the ones in the BD data.
The point $q_0$ where the corrections to $\beta$ scaling change sign can
be inferred from Fig.~\ref{beta} to be $q_0\approx9/d$. It is independent
on the type of short-time dynamics, in excellent agreement with the
predicted universality of structural relaxation, and in particular the
correction-to-scaling amplitudes. The numerical value of $q_0$ also agrees
well with that found in an analysis of the MCT results for the tagged-particle
correlator in a hard-sphere system \cite{Fuchs1998}, where the change occurs
at $q_{0,{\text{MCT}}}\approx9.3/d$.

The $\beta$ correlator for short times approaches the critical power law,
$G(t)\sim t^{-a}$. Comparing this asymptote with the fitted $\beta$
correlator in Fig.~\ref{beta}, one finds that the $t^{-a}$ law already
deviates from the $\beta$ correlator at $t\approx1$ for the BD data,
and at $t\approx0.3$ for the ND data. Thus the critical decay cannot be
identified from the simulation data. This is typical for most experimental
data \cite{Goetze1999}.
One thus has to be careful when extracting the exponent $a$
from an analysis of the $\beta$ relaxation.

Let us from now on restrict the discussion to the BD data set. For the ND
data, deviations from the MCT predictions occur in the early part of the
$\beta$ regime, and thus the theory can explain a larger part of the BD
curves than it can do for the ND ones. For the analysis of the long-time
universality of the dynamics outlined above, we conclude that these deviations
are not a feature of the glassy dynamics. It is known that MCT treats the
short-time dynamics insufficiently \cite{Kob2002}, and that Brownian
dynamics typically stays closer to the MCT scenario for a larger time window
than the corresponding Newtonian dynamics \cite{Franosch1998,Fuchs1999b}.

\section{Full MCT Analysis}\label{mct-analysis}

\begin{figure}
\includegraphics[width=.75\linewidth]{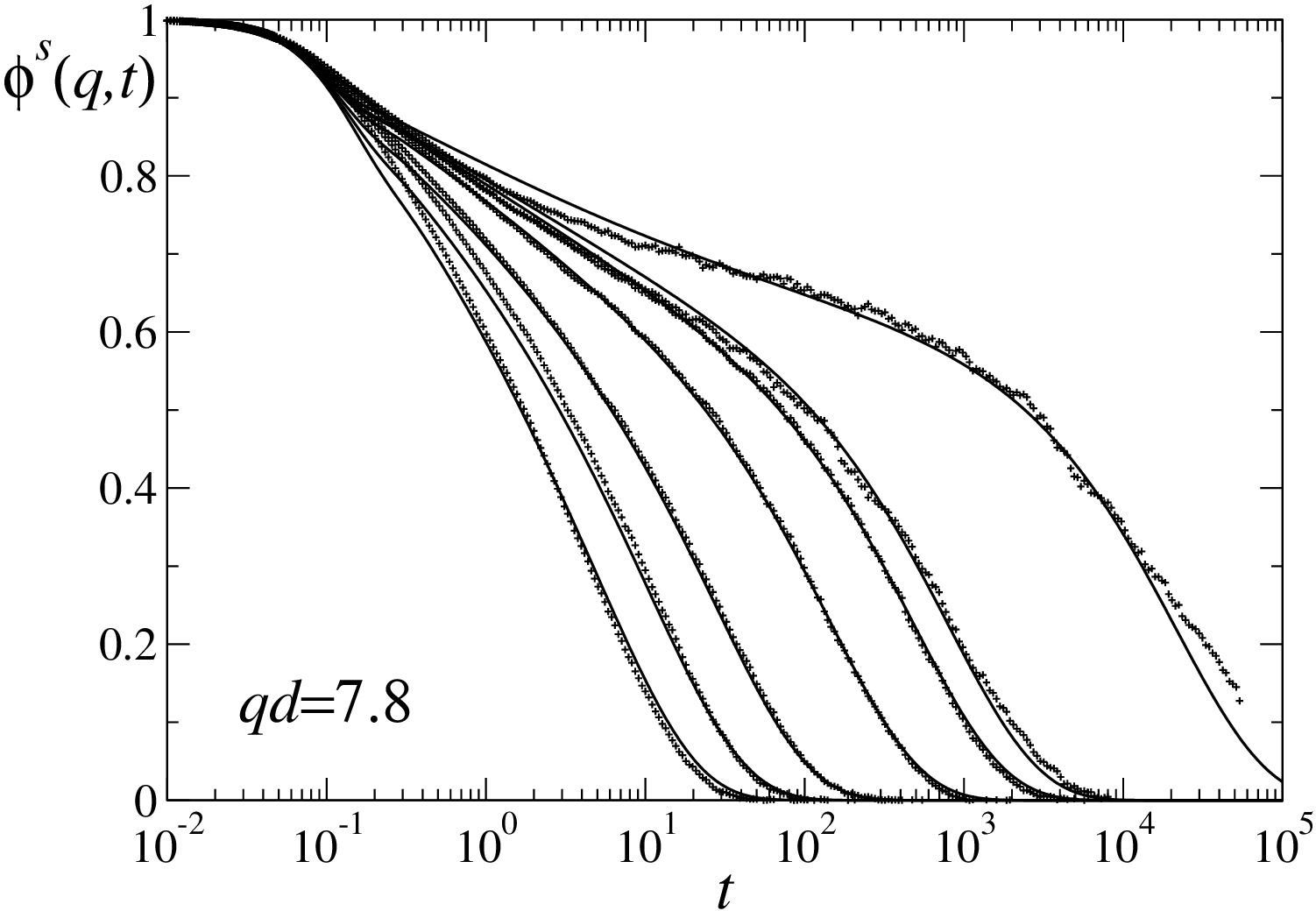}
\caption{\label{fit-q7}
  MCT (solid lines) and simulation results (symbols) for $\phi^s(q,t)$.
  For the simulation data, packing fractions are $\varphi=0.50$, $0.53$,
  $0.55$, $0.57$, $0.58$, $0.585$, and $0.59$ (from left to right), and
  $qd=7.8$. For the MCT results, packing fractions have been adjusted to
  $\varphi_{\text{MCT}}=0.445$, $0.47$, $0.484$, $0.499$, $0.505$, $0.508$, and
  $0.5135$, and the wave number has been adjusted to $q_{\text{MCT}} d=9.13$; see
  text for details.
}
\end{figure}

\begin{figure}
\includegraphics[width=.75\linewidth]{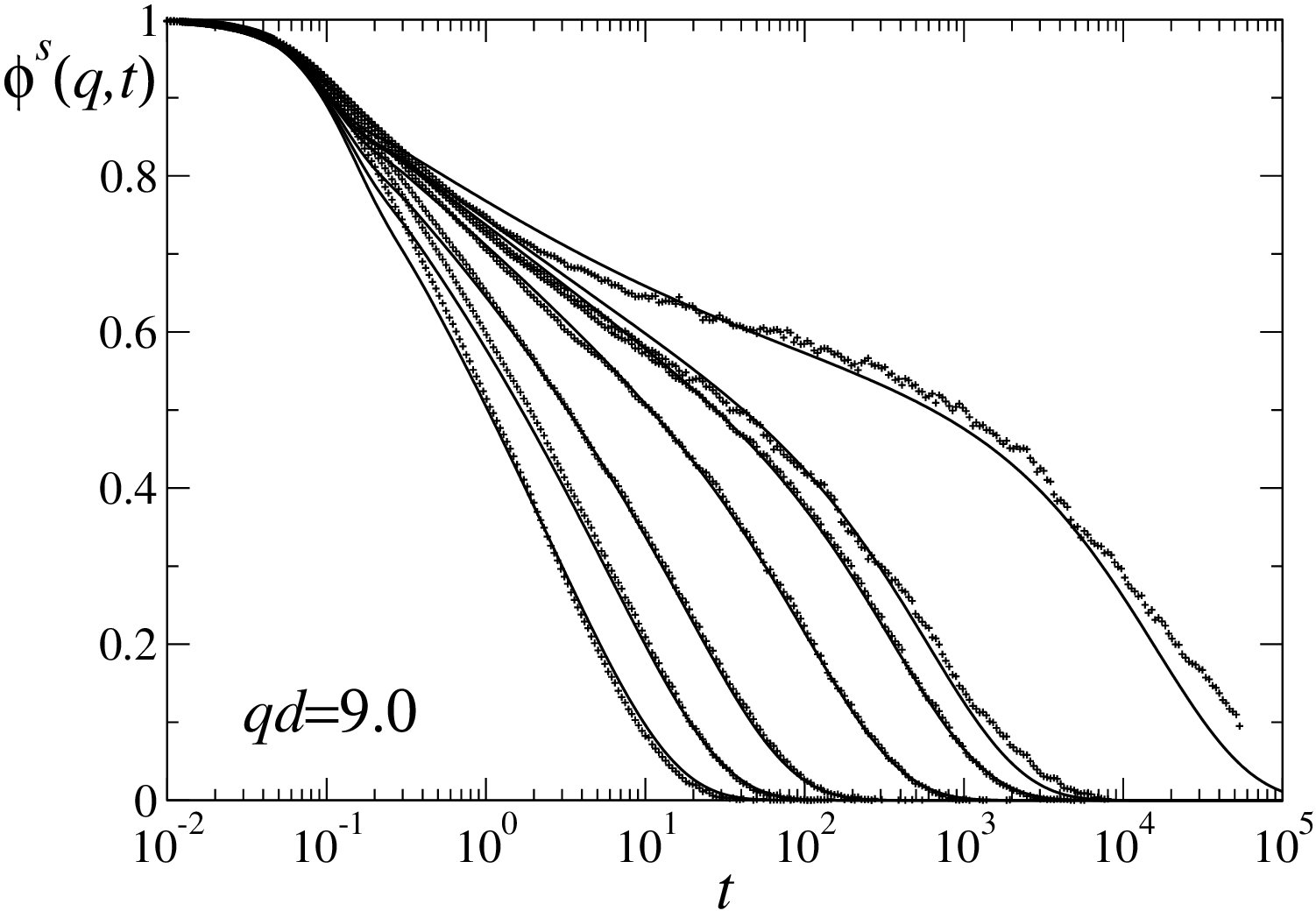}
\caption{\label{fit-q9}
  MCT and simulation data for $\phi^s(q,t)$ with symbols and packing
  fractions as in Fig.~\ref{fit-q7}, but for $qd=9.0$, adjusted
  in the MCT calculations to $q_{\text{MCT}} d=10.3$.
}
\end{figure}

\begin{figure}
\includegraphics[width=.75\linewidth]{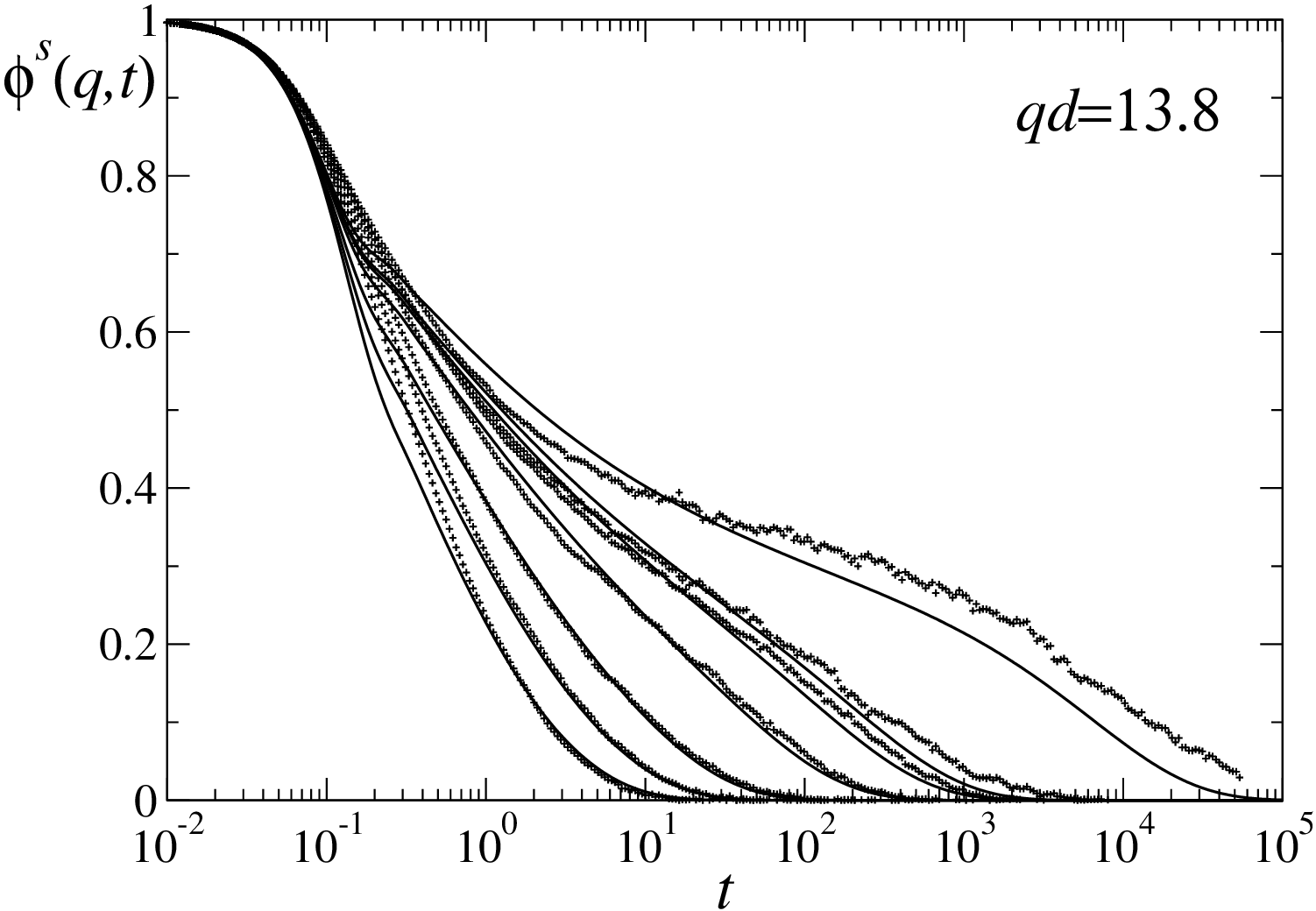}
\caption{\label{fit-q13}
  MCT and simulation data for $\phi^s(q,t)$ with symbols and packing
  fractions as in Fig.~\ref{fit-q7}, but for $qd=13.8$ ($q_{\text{MCT}} d=15.1$).
}
\end{figure}

\begin{figure}
\includegraphics[width=.75\linewidth]{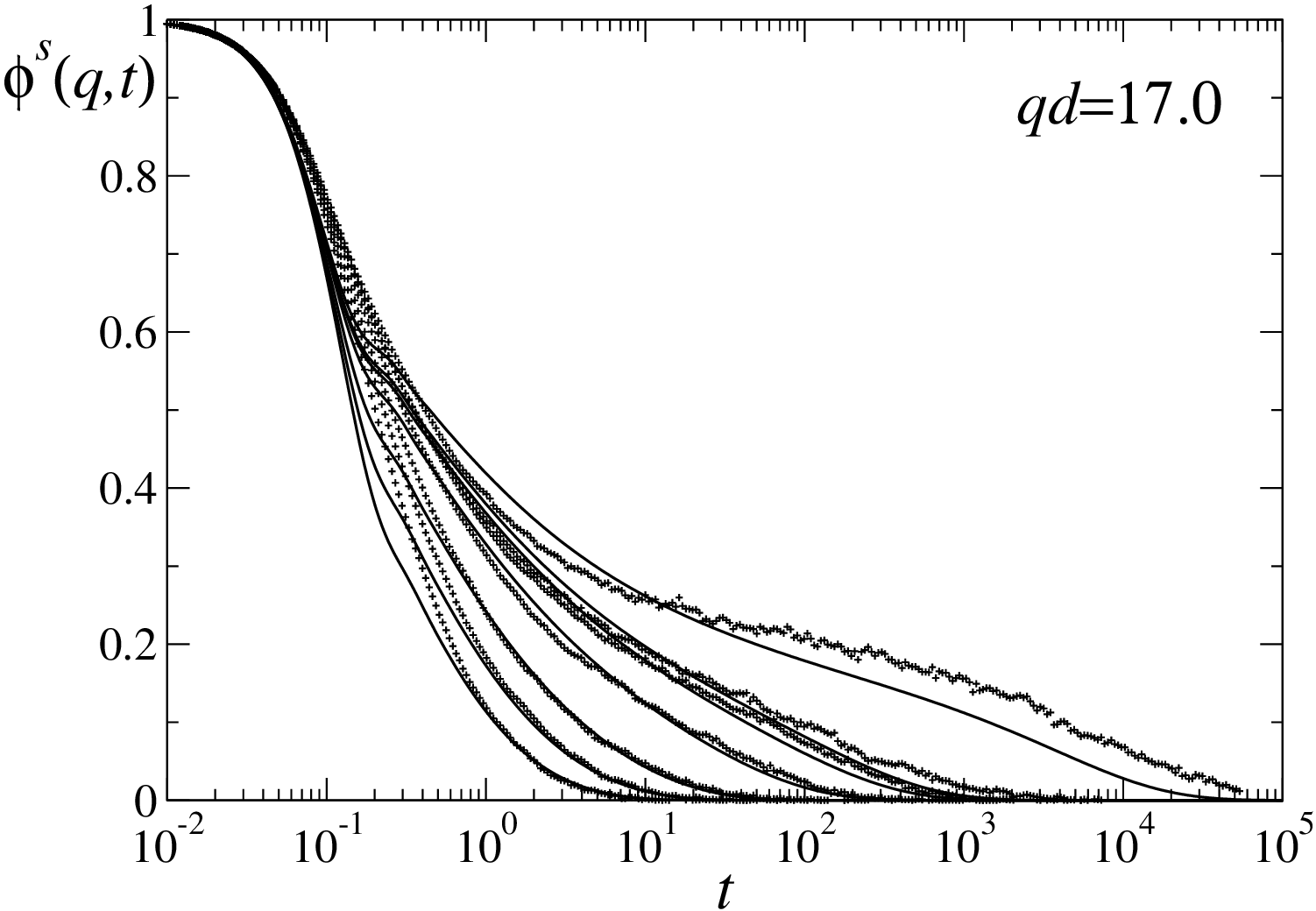}
\caption{\label{fit-q17}
  MCT and simulation data for $\phi^s(q,t)$ with symbols and packing
  fractions as in Fig.~\ref{fit-q7}, but for $qd=17.0$ ($q_{\text{MCT}} d=18.3$).
}
\end{figure}

We now turn to a full data analysis, i.e.\ a comparison of the complete
simulation data with the solutions of the full MCT equations for a hard-sphere
system.

In the MCT picture, the glassy dynamics of the hard-sphere system is
mainly driven by density fluctuations over the length scale of the mean
nearest-neighbour distance, i.e.\ with wave numbers close to that of the
first sharp diffraction peak in $S(q)$. We therefore begin the analysis
by focusing on the data for $q=7.8$. The results of our full-MCT fits
to the BD simulation data are shown in Fig.~\ref{fit-q7}. To achieve this and
the following fits, we have adjusted $\varphi_{\text{MCT}}$ for each curve and allowed
the wave number $q_{\text{MCT}}$ to vary slightly with respect to the correct value
$q$. No other parameters have been adjusted. As noted above, the $q$-shift
to some extent accounts for a mismatch in length scales between the simulation
and the theory predictions. The adjustment of $\varphi_{\text{MCT}}$
on the other hand accounts for the known error in $\varphi^c$. We will discuss
the relation of the fitted $\varphi_{\text{MCT}}$ to the correct packing fraction
$\varphi$ below.

The fit shown in Fig.~\ref{fit-q7} demonstrates that the theory can, with
these modifications allowed, account
for the dynamics of the hard-sphere system in a time window of over four
decades on a $10\%$ error level. Only at larger times, $t\approx10^4$ in
our units, i.e.\ at the highest packing fraction studied, systematic
deviations are observed. The simulation for this packing fraction shows slower
dynamics than expected from the theory. Also the shape of the final decay is
different, as noted above in connection with $\alpha$ scaling.
On the short-time side, the MCT description works down to a time $t\approx1$.
At shorter times, it is still almost quantitative, but one observes a different
curve shape. The simulation data appears more strongly damped than the MCT
curves. This is to be expected, since neglecting the regular part
of the memory kernel in Eq.~\eqref{mctmemreg} will lead to such deviations
at short times. We could have accounted for this partly by choosing a
higher value of $\nu$ in Eq.~\eqref{mctmemreg}, but we have not done so since
we are not concerned with the short-time dynamics in this work.

Once the fit for $q=7.8$ was completed to fix the empirical relation
$\varphi_{\text{MCT}}(\varphi)$, we have
analyzed data for other wave numbers up to $q=30$, hereby fixing the
relation $q_{\text{MCT}}(q)$. Typical results for
$q\le17$ are exhibited in Figs.~\ref{fit-q9} to \ref{fit-q17}.
Note that the only parameter that was adjusted for these fits is the wave
number, $q_{\text{MCT}}$. This way, Figs.~\ref{fit-q9} to \ref{fit-q17} demonstrate
how MCT is able to reproduce the wave-vector dependent changes in the
structural-relaxation window. At even higher $q$, it becomes too difficult to
judge the fit quality, since the $f(q)$ are close to zero there.
Connected with this is the growing influence of the microscopic regime,
$t\le1$, on the main part of the decay of the correlators with increasing $q$.
For $q=17$ and the highest packing fractions, already about $60\%$ of the
decay of $\phi^s(q,t)$ from unity to zero are made up of such microscopic
dynamics. Consequently, the errors made in its description are to be seen
more explicitly in Fig.~\ref{fit-q17} than in Fig.~\ref{fit-q7}.

Apart from this, also the fits at $q>7.8$ using the full-MCT results are
quite satisfactory in the time window $1\le t\le10^4$, save the highest
simulated density, for which errors are largest and extend down to $t\approx10$
for $q=13.8$ and $17$, cf.\ Figs.~\ref{fit-q13} and \ref{fit-q17}.
One notices a trend that the
$\alpha$-relaxation dynamics becomes slower in the simulation data than expected
from the MCT fits, i.e.\ the local dynamics is slower than one
estimates in the theory. The finding can probably not
fully be attributed to the incorrect structure-factor input used, since
a recent study of binary hard-sphere mixtures reported a similar discrepancy
for the $q$-dependence of the $\alpha$-relaxation times even when basing
MCT on the `correct' simulated $S(q)$ as input \cite{Foffi2004}. The same
trend is also apparent in the full-MCT analysis of a binary Lennard-Jones
mixture \cite{Kob2002}.

\begin{figure}
\includegraphics[width=.75\linewidth]{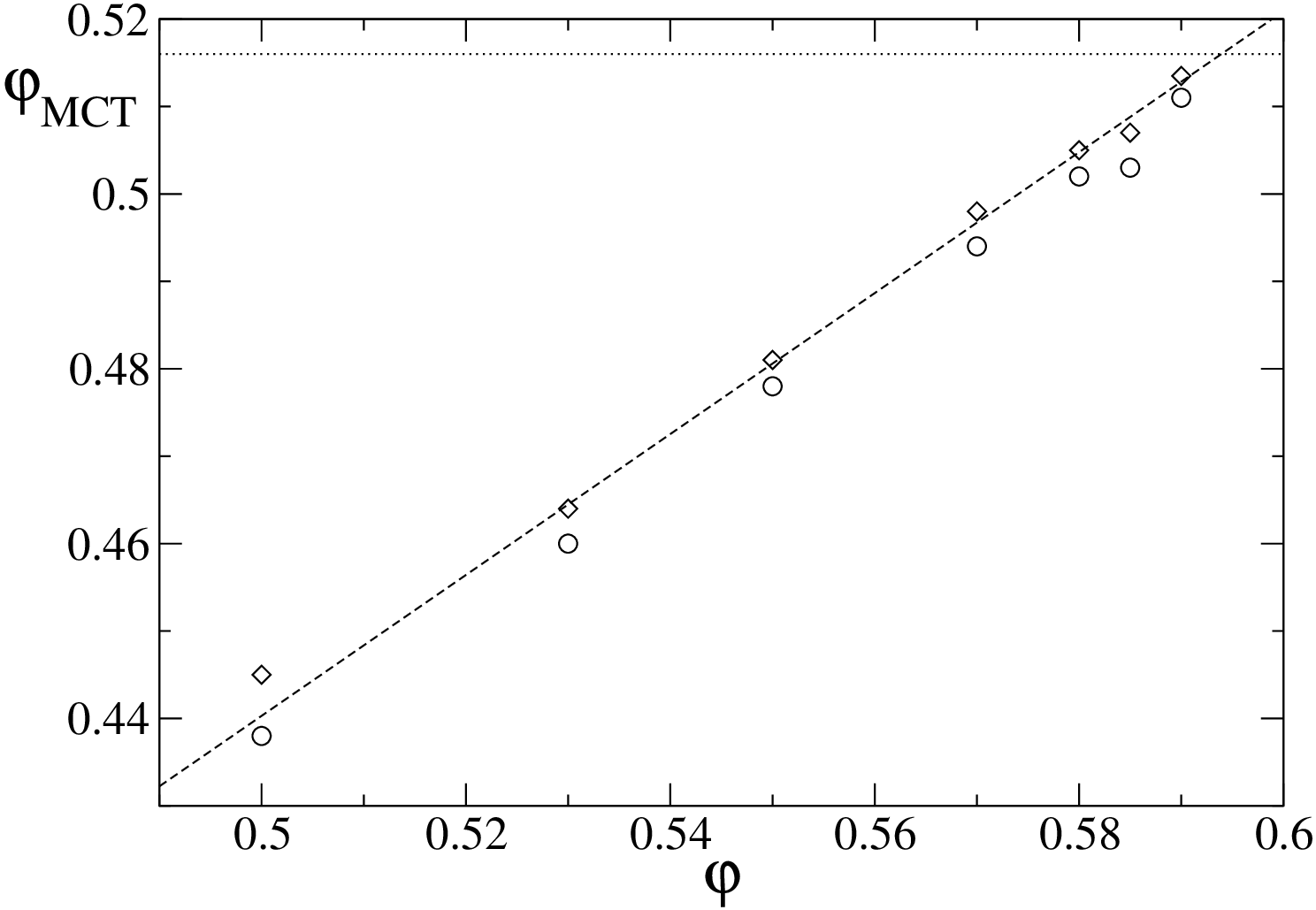}
\caption{\label{phimap}
  Plot of $\varphi_{\text{MCT}}$ vs.\ $\varphi$ for the fits shown in
  Figs.~\protect\ref{fit-q7}--\protect\ref{fit-q17}, \protect\ref{fit-q4} and
  \protect\ref{fit-msd} (diamonds). The dashed line is a linear fit,
  $\varphi_{\text{MCT}}\approx 0.81\varphi+0.037$. The circles are for the independent
  fit of the MSD, Fig.~\protect\ref{fit-msd-extra}.
  The dotted horizontal line indicates the calculated critical point,
  $\varphi^c\approx0.516$.
}
\end{figure}

Having established the overall quality of the MCT description for the
structural dynamics on length scales smaller and comparable to the typical
particle-particle distance,
let us now investigate the adjustment in $\varphi_{\text{MCT}}$
needed to achieve this level of agreement. A plot of $\varphi_{\text{MCT}}$ vs.\
$\varphi$ is shown in Fig.~\ref{phimap} (diamond symbols). The figure
reveals that the relation is close to linear, and thus the nontrivial
variation of the relaxation curves close to the singularity $\varphi^c$
has not been put in `by hand' through the fitting parameter $\varphi_{\text{MCT}}$.
We can estimate the correct value of the glass-transition packing
fraction by a linear fit to the $\varphi_{\text{MCT}}$-versus-$\varphi$ curve.
Using the calculated value $\varphi^c_{\text{MCT}}\approx0.516$, we get
$\varphi^c\approx0.594$. This value is nicely consistent with the one
obtained from the $\alpha$-scale analysis of the data,
cf.\ Fig.~\ref{taualpha}, and also from an earlier analysis of the
simulations \cite{Puertas2002}.
Note that it differs from the result obtained from MCT based on the
simulated $S(q)$, $\varphi^c\approx0.585$ by less than $2\%$.
The slope of the linear fit in Fig.~\ref{phimap} is not equal to unity, and
its zero is shifted. If one considers the connection of the distance parameter
$\sigma$ of MCT to the control-parameter distance $\varepsilon$, this
translates into an error of the leading-order constant of proportionality $C$
in Eq.~\eqref{sigmaequation}.
From Fig.~\ref{phimap} we conclude that the value calculated from MCT,
$C_{\text{MCT}}\approx1.54$ \cite{Franosch1997},
is in error by about $20\%$, $C\approx1.2$.

\begin{figure}
\includegraphics[width=.75\linewidth]{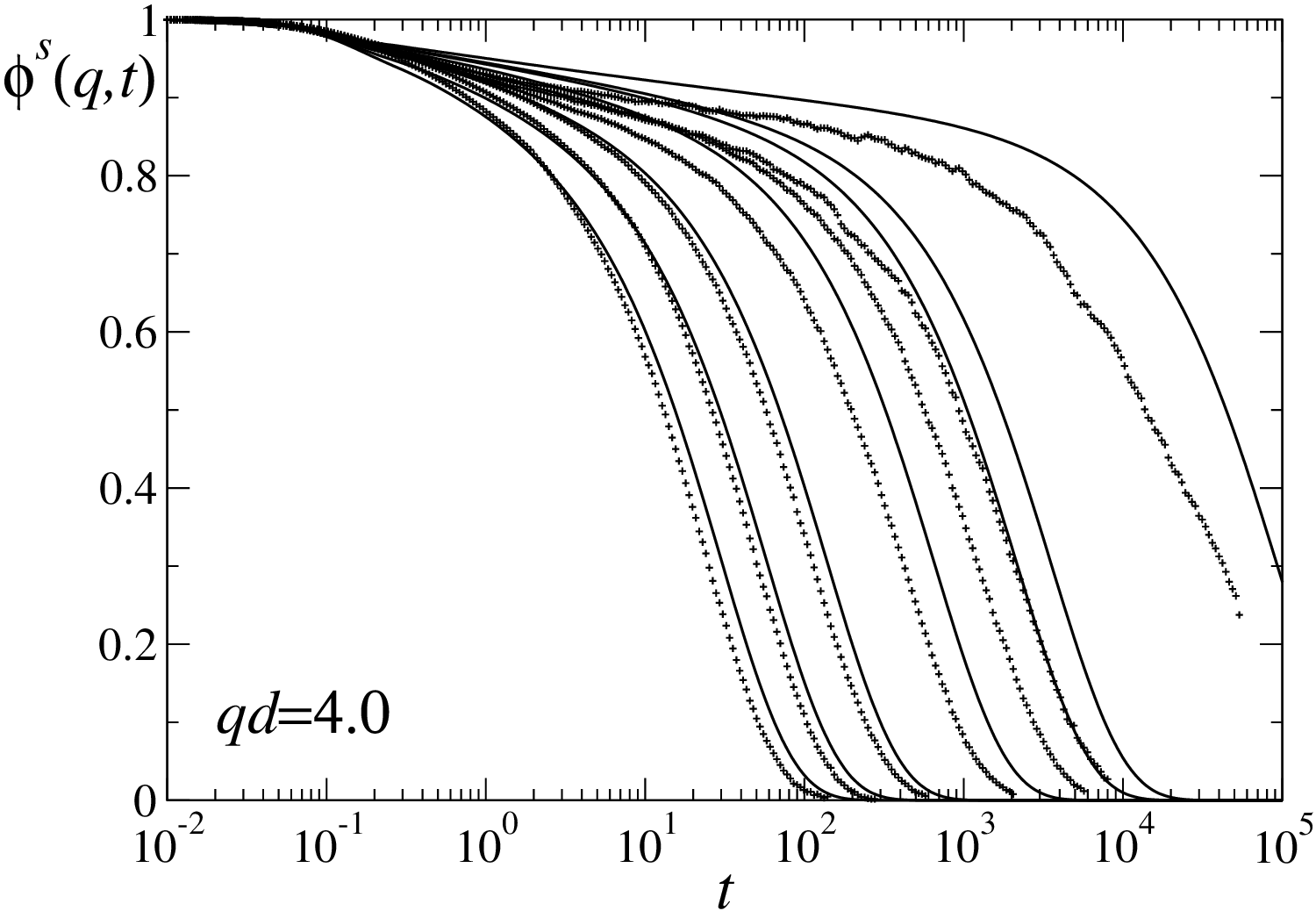}
\caption{\label{fit-q4}
  MCT and simulation data for $\phi^s(q,t)$ with symbols and packing
  fractions as in Fig.~\ref{fit-q7}, but for $qd=4.0$ ($q_{\text{MCT}} d=5.0$).
}
\end{figure}

For small $q$, the MCT description of the data shows larger quantitative
errors, while it remains qualitatively correct. This is exhibited by the
fits done for $q=4.0$, Fig.~\ref{fit-q4}. Again, only $q_{\text{MCT}}$ was allowed
to vary, while the $\varphi_{\text{MCT}}$ have been determined from the
$q=7.8$ fit shown in Fig.~\ref{fit-q7}. While for this latter fit, the
MCT fits reproduce the $\alpha$-relaxation times rather well, this is not
the case for the $q=4.0$ fit at $\varphi\ge0.57$. Instead, one observes
a systematic trend for the simulation data to decay increasingly faster than
the MCT curves with increasing $\varphi$. In addition, the deviations observed
in the $\beta$-relaxation window, while still within a $10\%$ level, are
larger for $q=4.0$ than they are for $q\ge7.8$. Even more, it was
necessary to allow for a $25\%$ deviations between $q_{\text{MCT}}$ and $q$,
whereas this deviation was less than $17\%$ for all other fits.
This last finding suggests that the $f^{s,c}(q)$-versus-$q$ curve calculated
within MCT is too broad. It is common to express deviations of the
$\phi^{s,c}(q,t)$-versus-$q$ curve at fixed $t$ from a Gaussian at small $q$
in terms of the non-Gaussian parameter, $\alpha_2(t)$. These non-Gaussian
corrections are reproduced qualitatively correct by MCT, but with an error
in magnitude. One finds $\alpha_2(t)\le0.3$ in the
theory \cite{Fuchs1998}, while for our simulation,
$\alpha_2(t)$ reaches values up to $2.5$ in both BD and ND, which is in
agreement with similar simulation results for other systems
\cite{Kob1995}.
But note that for times where
$\phi^s(q,t)$ is close to its plateau value $f^{s,c}(q)$, both the MCT and
the simulation value of $\alpha_2(t)$ are positive. Thus the underestimation
of $\alpha_2$ in the theory would let the $f^{s,c}(q)$-versus-$q$ curve
appear too narrow, opposite to what is observed from our fits. We thus
conclude that non-Gaussian corrections as expressed through $\alpha_2(t)$
and the quantitative error MCT makes in expressing them
cannot be alluded to to explain the deviations observed at $q=4.0$.
Let us point out that the deviations discussed above do not depend
significantly on the fact that we have based the MCT calculation on the
PY-DCF. Using the simulated structure factor with MCT gives correlation
functions $\phi^s(q,t)$ that behave qualitatively as the ones shown here.

\begin{figure}
\includegraphics[width=.75\linewidth]{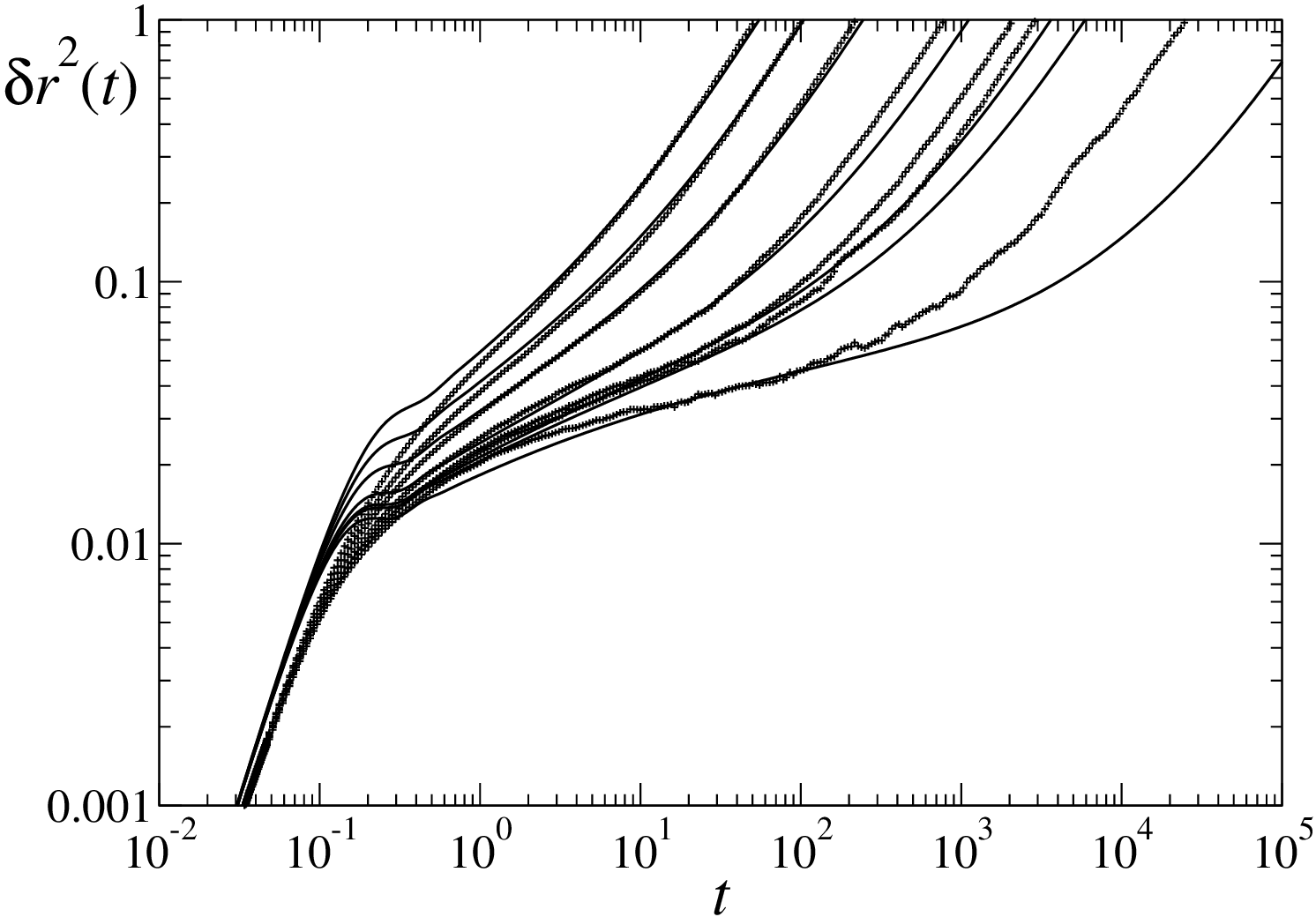}
\caption{\label{fit-msd}
  Comparison of the mean-squared displacements $\delta r^2(t)$ from simulation
  and MCT calculations; all fit parameters have been taken from
  Fig.~\ref{fit-q7}. In addition, the MCT curves have been multiplied by
  $1.1$ to account for an error in localization length; see text for details.
}
\end{figure}

It is instructive to extend this analysis towards the mean-squared displacement
data. Since the MSD is given through a memory kernel that basically is
a $q\to0$ limit of the tagged-particle-correlator memory kernel,
Eq.~\eqref{msdequation}, its analysis can be viewed as the $q\to0$ extension of
the above fitting procedure.

We report the BD simulation data for the MSD, together with the MCT curves
according to the correlation functions shown above, in Fig.~\ref{fit-msd}.
For the MSD, no wave number is to be adjusted, and in this sense, the MCT
results of Fig.~\ref{fit-msd} are not fitting results, but rather consequences
of the analysis done for $q=7.8$, shown in Fig.~\ref{fit-q7}. We have,
however, adjusted a global length scale in this plot, in order to better fit
the localization plateau visible in the data. The MCT curves have been scaled
up by a factor of $1.1$, which accounts for a $5\%$ underestimation of the
localization length by the theory.
Note that at short times, $t<1$, the description of the data using MCT is
of similar quality as discussed above. In particular, the MSD plot reveals
that the BD simulation still resembles a Newtonian short-time dynamics,
though strongly damped: the MSD roughly follows a $\delta r^2\sim t^2$
law for $10^{-2}<t<10^{-1}$, and not a $\delta r^2\sim t$ law as would be
expected for short-time diffusion in a strictly Brownian system. Theory and
simulation do not match at short times because of the scale factor applied.
For long times, a qualitatively similar picture emerges as for $q=4.0$,
regarding the variation of the $\alpha$-relaxation times with $\varphi$,
now showing through a displacement of the long-time diffusive straight
lines in the $\delta r^2(t)$ plot. As for $q=4.0$, the quality of the MCT
description of the $\beta$-relaxation regime similarly is worse than for
$q\ge7.8$. But for the MSD, all deviations are larger than for $q=4.0$,
especially in the $\alpha$-relaxation regime.

\begin{figure}
\includegraphics[width=.75\linewidth]{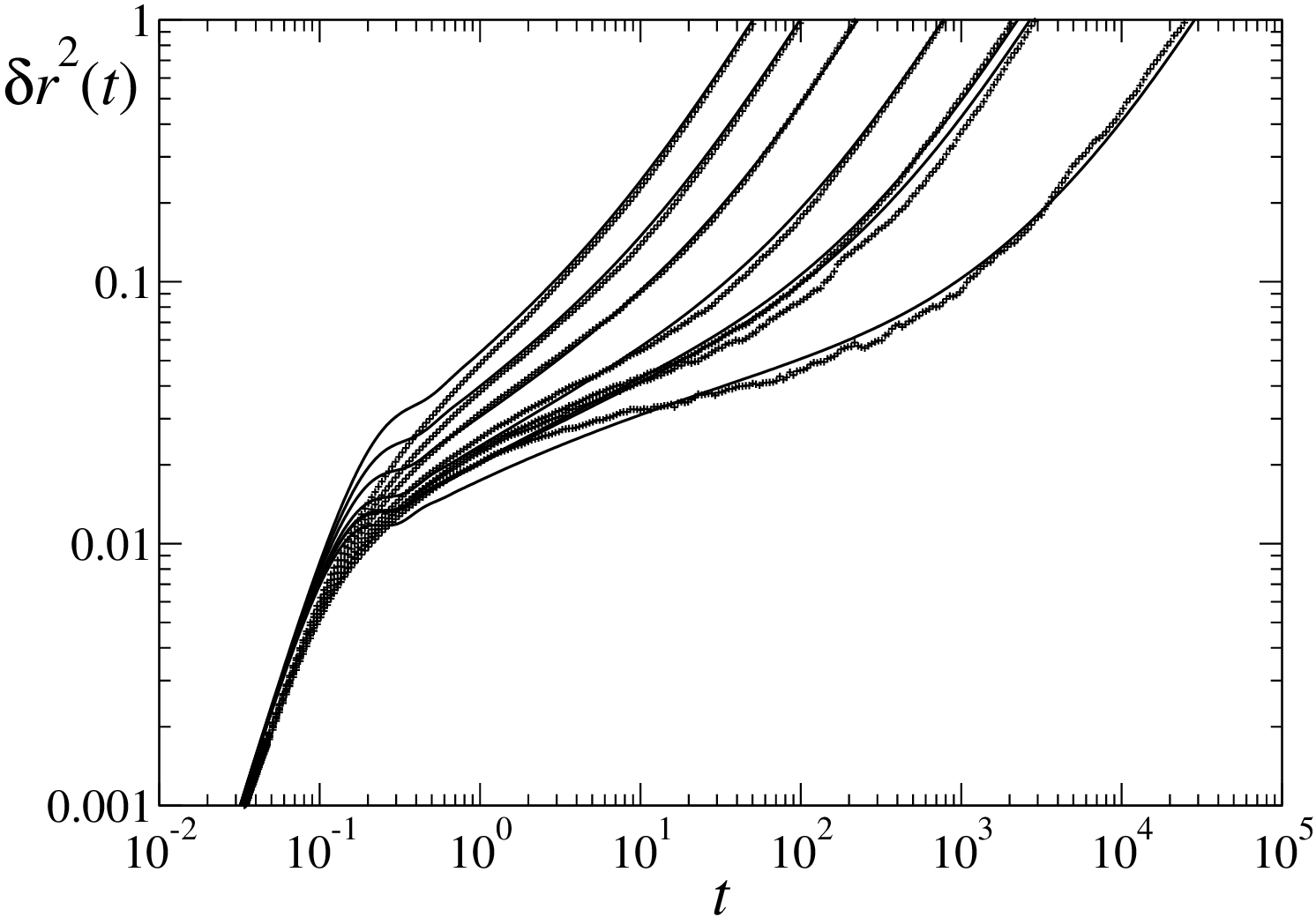}
\caption{\label{fit-msd-extra}
  Comparison of the mean-squared displacements $\delta r^2(t)$ from simulation
  and MCT calculations. In this plot, values $\varphi_{\text{MCT}}$ used for the
  MCT calculations have been adjusted to fit the long-time diffusion regime
  of the simulated $\delta r^2(t)$; the values are $\varphi_{\text{MCT}}=0.438$,
  $0.46$, $0.478$, $0.494$, $0.502$, $0.503$, and $0.511$ for
  $\varphi=0.50$, $0.53$, $0.55$, $0.57,$ $0.58$, $0.585$, and $0.59$,
  respectively.
}
\end{figure}

Before investigating the $\alpha$-relaxation regime in more detail,
let us note that, all deviations taken aside, the shapes of the MSD
curves as predicted by MCT are qualitatively correct. To substantiate this
statement, Fig.~\ref{fit-msd-extra} shows an independent fit of the MSD using
MCT. Instead of fixing the $\varphi_{\text{MCT}}$-versus-$\varphi$ relation from
the data at $q=7.8$, as was done above, in this case this relation was
determined from the MSD alone. It is noteworthy that by correcting the error
in the $\alpha$-relaxation time scale observed before, also the description
of the $\beta$-relaxation window improves. In particular, we did not scale
the MCT results as we have done in Fig.~\ref{fit-msd}. The values
$\varphi_{\text{MCT}}(\varphi)$ used in Fig.~\ref{fit-msd-extra} are reported
in Fig.~\ref{phimap} as the circle symbols. They also lie on a straight line,
which is shifted downward somewhat with respect to the original
relation deduced from the above fits. In turn, an estimation of $\varphi^c$
from the mean-squared displacement, i.e.\ from the diffusivities, yields
a value that is too high, viz.: $\varphi^c\approx0.598\pm0.003$. This is
a typical result also found in other simulations \cite{Foffi2004},
but not in accord with MCT. But note that the estimations for $\varphi^c$
are quite close to each other, so the deviations can be regarded as
indications of error margins.
As a side remark, let us note that the parameters from the independent fit to
the MSD presented in Fig.~\ref{fit-msd-extra} could be used to improve the
description of the $q=4.0$ case somewhat, but not completely. One concludes
that the MCT description of the single-particle
structural relaxation smoothly deteriorates for decreasing $q$.

\begin{figure}
\includegraphics[width=.75\linewidth]{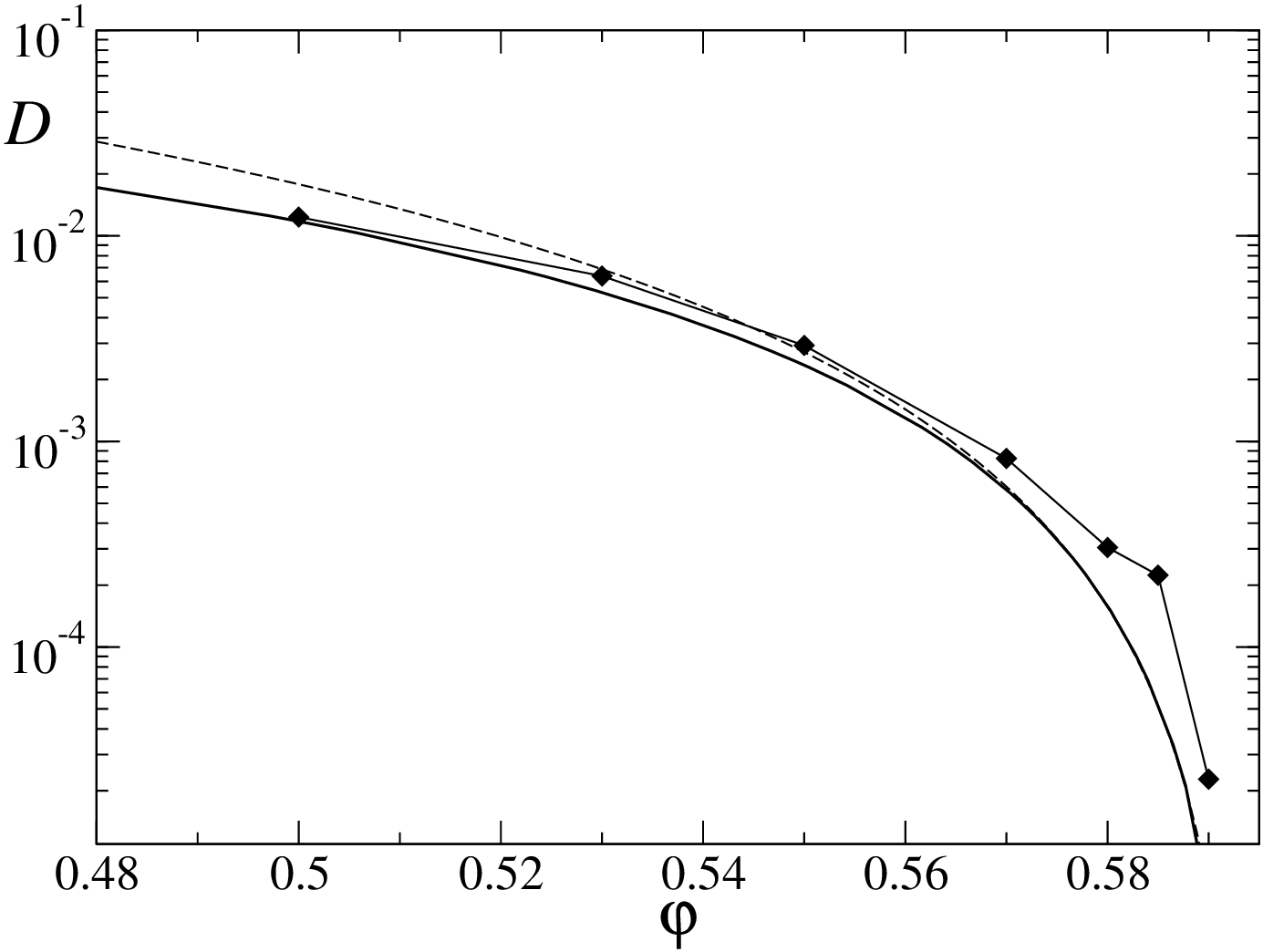}
\caption{\label{diffusionplot}
  Self-diffusion coefficients $D$ as a function of the packing fraction
  $\varphi$, as obtained from the Brownian dynamics simulation (diamonds,
  with connecting lines to guide the eye) and from MCT calculations
  (solid line). The dashed line indicates the MCT asymptote,
  $D\propto (\varphi^c-\varphi)^\gamma$.
}
\end{figure}

The deviations in the $\alpha$-relaxation regime, i.e.\ the long-time
diffusive regime, that arise for the MSD can be analyzed more clearly by
looking at the long-time self-diffusion coefficient $D(\varphi)$ itself.
Here, $D$ has been determined from the simulations by the Einstein relation,
$\delta r^2\sim 6Dt$ for large $t$, at times $t$ where $\delta r^2$ is of
the order of $10$ squared particle radii. The results for the BD simulations are
shown in Fig.~\ref{diffusionplot} as the diamond symbols. In contrast,
the MCT calculation, plotted in the figure as a solid line, shifted according
to the relation $\varphi_{\text{MCT}}(\varphi)$ used in the above discussion,
systematically falls below these values. The relative error in $D$ is less
than $20\%$ up to $\varphi\approx0.55$, increases to $80\%$ at
$\varphi=0.58$, and reaches a factor of $4$ at $\varphi=0.59$.
The two curves could be matched within the error bars by further shifting
the MCT results along the $\varphi$ axis by less than $1\%$, which is
basically what has been done in Fig.~\ref{fit-msd-extra}. But let us stress
that there is no theoretical justification for doing so.

MCT predicts a power-law asymptote for $D$ with the same exponent $\gamma$
that applies for the divergence of the $\alpha$-relaxation times,
$D\propto|\sigma|^\gamma$. This asymptote is included in
Fig.~\ref{diffusionplot} for the MCT results as a dashed line. It is also
possible to fit the simulation data with such a power-law. We have restricted
these fits to $\varphi\ge0.55$ and have omitted the value at $\varphi=0.585$.
If we fix the exponent to the theoretical value, $\gamma\approx2.46$, we get
reasonable agreement with a fitted $\varphi^c\approx0.599$, in agreement with
the above observation. If we, on the other hand, also determine $\gamma$ from
the fit to $D(\varphi)$, the result is $\varphi^c\approx0.597$ with
$\gamma\approx2.24$. It is a typical observation from simulation and
experiment is that an independent determination of $\gamma$ from the diffusion
coefficient yields a different value than that obtained from the analysis of
the density correlators \cite{Kob1994}. From the comparison of the MCT
results with the asymptotic prediction in Fig.~\ref{diffusionplot}, it is,
however, clear that the asymptotic law only holds for $D<10^{-3}$, i.e.\
for $\varphi\ge0.58$ for our simulations. Thus a large part of the fitted
simulation data is outside the asymptotic regime for this power law,
and the fit yields an effective exponent rather than the true $\gamma$.

\begin{figure}
\includegraphics[width=.75\linewidth]{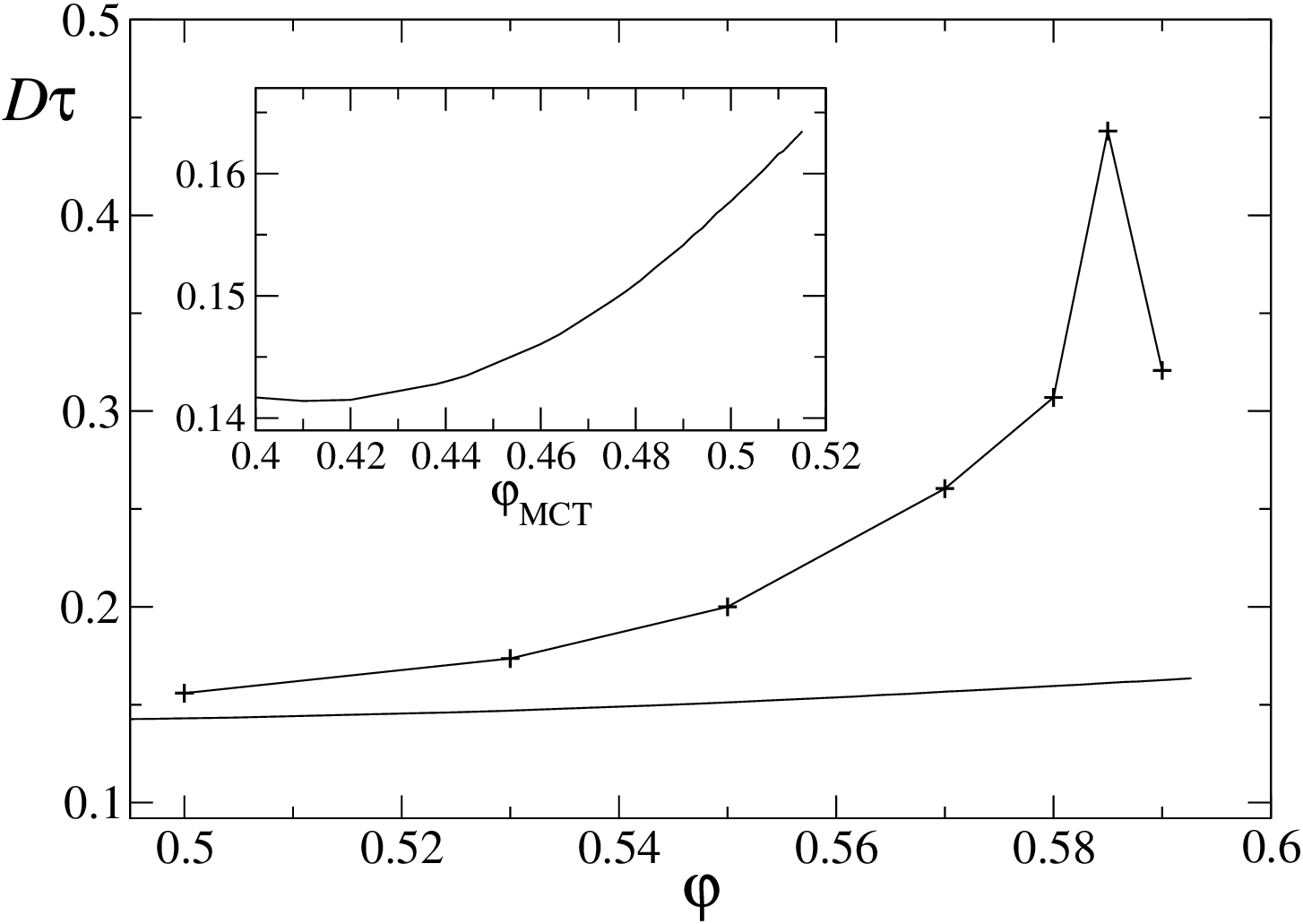}
\caption{\label{dtau}
  Product $D\cdot\tau(q)$ at wave vector $qd=7.8$ for the BD simulation
  (crosses, connected by lines to guide the eye), and for the MCT curves
  (solid line). The latter curve has been transformed along the horizontal
  axis according to Fig.~\ref{phimap}.
  The inset shows a magnification of the MCT result versus $\varphi_{\text{MCT}}$.
}
\end{figure}

The above results indicate that with $\varphi$ increasing close to
$\varphi^c$, a decoupling of the diffusion time scale, as seen from the
mean-squared displacement, from the density-fluctuation-relaxation time scale,
as seen in the density correlation functions,
sets in. This can be illustrated without referring to any fits
by plotting the product $D\tau(q)$ of the diffusion coefficient $D$ and
the $\alpha$-relaxation time scale \cite{Kob2003}.
For the latter, let us choose the value
obtained for $q=7.8$, as a typical representative of the local-order
length scale. Figure~\ref{dtau} shows as symbols the results from the
BD simulation.
One notices an increase in $D\tau$
by a factor of $2$ to $3$ within the density range covered by the simulations.
We have also checked that this holds similarly for $q=4.0$ and $q=9.0$.
The corresponding MCT result is shown as the solid line in Fig.~\ref{dtau},
which is magnified in the inset of the figure.
Here, the product $D\tau$ also increases with increasing $\varphi$ close to
$\varphi^c$, but only on the order of $10\%$. One thus concludes that there
is a rather large quantitative error in this quantity, although not
necessarily a qualitative one. MCT predicts that $D\tau$ approaches a finite
value as $\varphi\to\varphi^c$. As to whether $D\tau$ diverges or stays
finite at $\varphi^c$ in the simulation, our data remain inconclusive.
Note that the values for the highest two packing fractions simulated are
relatively uncertain, as the scatter in $D\tau$ indicates.

\section{Polydispersity Effects}\label{polydisp}

Up to now, we have neglected the fact that the simulated system is not
strictly a single-component system. Instead, some size
polydispersity was needed in order to avoid crystallization.
In this section, we give a brief account of how much we expect this
small polydispersity to affect the results discussed above.

\begin{figure}
\includegraphics[width=.75\linewidth]{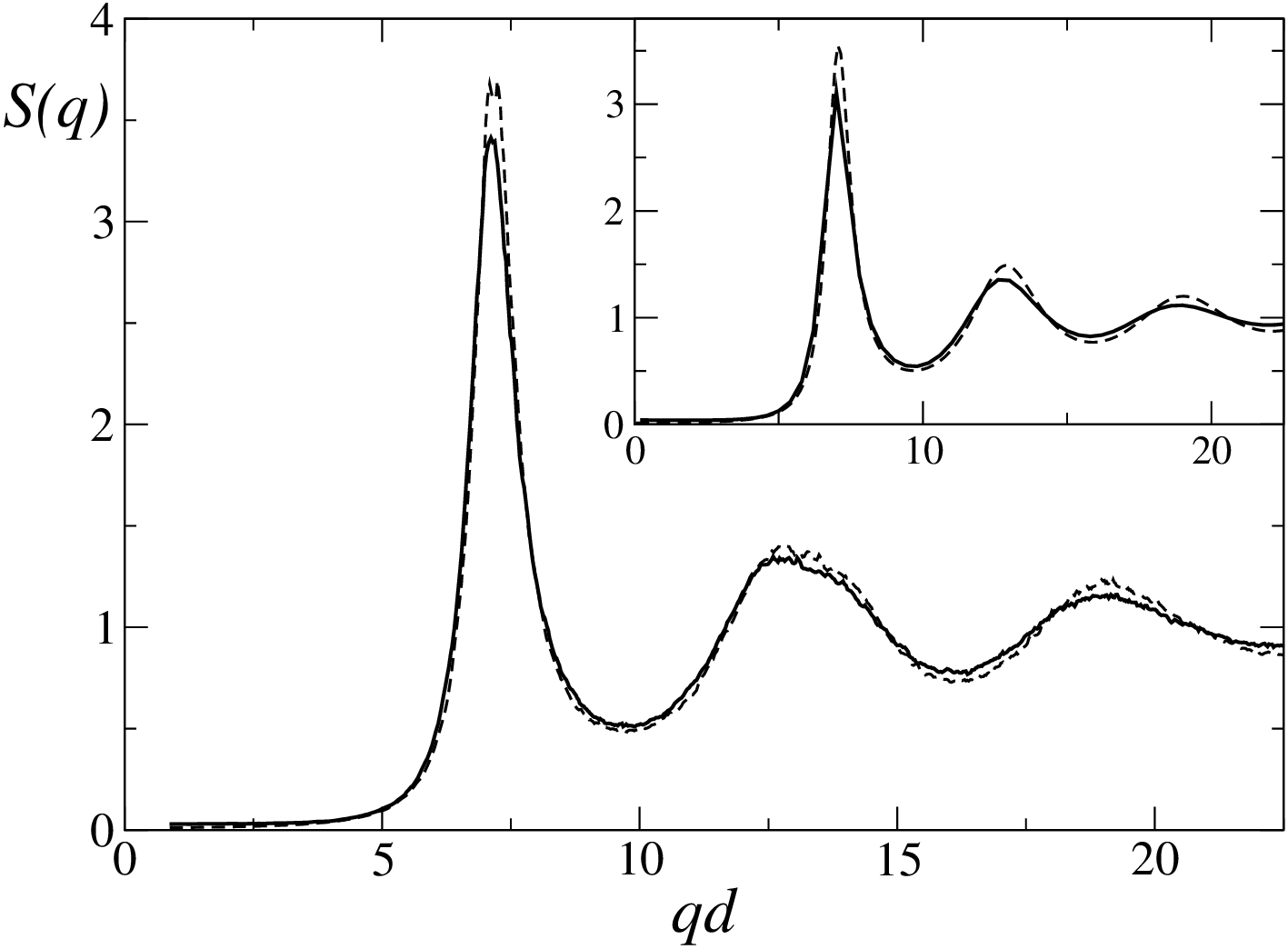}
\caption{\label{sq-pd-compare}
  Averaged static structure factor $S(q)$ for a monodisperse system
  (dashed lines) and the polydisperse system studied in this work
  (solid lines) at the same state point. The main figure shows the
  simulation results at $\varphi=0.54$. The inset shows the Percus-Yevick
  $S(q)$ at the respective MCT-critical packing fractions $\varphi_{\text{MCT}}^c$,
  for both the monodisperse (dashed) and a five-component system (solid
  lines).
}
\end{figure}

We first examine the influence of polydispersity on the equilibrium fluid
structure. To this end, we have simulated a monodisperse system of the same
soft spheres as used in the polydisperse simulations.
We found such simulations possible for packing fractions up to
$\varphi\approx0.54$, above which crystallization as monitored by $Q_6$
sets in rather quickly. The static structure factor $S(q)$ at this state point
is compared to the one from the polydisperse system at the same density
in Fig.~\ref{sq-pd-compare}. As expected, the polydisperse system exhibits
less pronounced ordering, visible in reduced oscillation amplitudes in
$S(q)$. The effect is well explained by the PY approximation, as the inset
of Fig.~\ref{sq-pd-compare} demonstrates. There, the (total-density) structure
factor for the one-component hard-sphere system is compared with that obtained
from the five-component mixture introducted in Sec.~\ref{mctdetails}.
One might expect the visible
differences in the monodisperse and the polydisperse $S(q)$, however small, to
affect the MCT results for $\varphi^c$. This would
be true if both systems were treated as one-component systems. But it is not
necessarily true for a full calculation of multi-component MCT, using the
full matrix of partial structure factors instead of only the averaged $S(q)$,
as we will discuss now.

Let us compare the results obtained for $f^{s,c}(q)$ for the
one-component system with those of the five-component system
mentioned above. This comparison is included in
Fig.~\ref{fcomp}, where the dashed line
shows the averaged $f^{s,c}_{\text{pd}}(q)$ according to
Eq.~\eqref{ncompavg}. The thus obtained curve is slightly narrower than the
solid curve, representing the result of the one-component calculation.
Accordingly, the average localization length increases slightly, by about
$4\%$. From the above discussion we conclude that this is a change in the
right direction, but not enough to account for the wave-number shift we
had to apply to describe the density correlation functions with the
one-component system.
Comparing with the Kohlrausch amplitudes $A(q)$ also shown in
Fig.~\ref{fcomp}, we note that the intrinsic error in determining the
plateau values from the data is larger than the differences between the two
MCT curves.

The values of $\varphi^c$ obtained from the MCT calculations with
one, three, and five components show only minor differences. While the
one-component result is $\varphi^c\approx0.5159$, we get $\varphi^c_{S=3}
\approx0.5153$, and $\varphi^c_{S=5}\approx0.5154$. Similarly, the exponent
parameter only changes slightly between these systems: from
$\lambda\approx0.735$ in the one-component system to
$\lambda_{S=5}\approx0.737$ for the five-component case. These changes in
$\varphi^c$ and $\lambda$ are significantly smaller than the uncertainty
in these quantities coming from the approximation used for the
static structure factor. Note that the value of $\varphi^c$ decreases
slightly in the multi-component systems mentioned. This is consistent with
recent MCT predictions for a two-component system \cite{Goetze2003}, where
it was found that for size ratios $d_{\text{small}}/d_{\text{large}}\ge0.8$
the critical point $\varphi^c$ slightly decreases compared with the
one-component system. Only when the size ratio became more extreme,
$d_{\text{small}}/d_{\text{large}}\le0.6$, say, did the MCT calculations
show a notable effect on $\varphi^c$. In this latter case, the values of
$\varphi^c$ were found to be larger in the mixture than in the one-component
system. This increase is commonly expected for polydisperse systems.
But from our calculations we conclude that such polydispersity-induced
fluidization does not occur for the present polydisperse size distribution,
which in particular lacks any large- or small-radius `tails'.

\begin{figure}
\includegraphics[width=.75\linewidth]{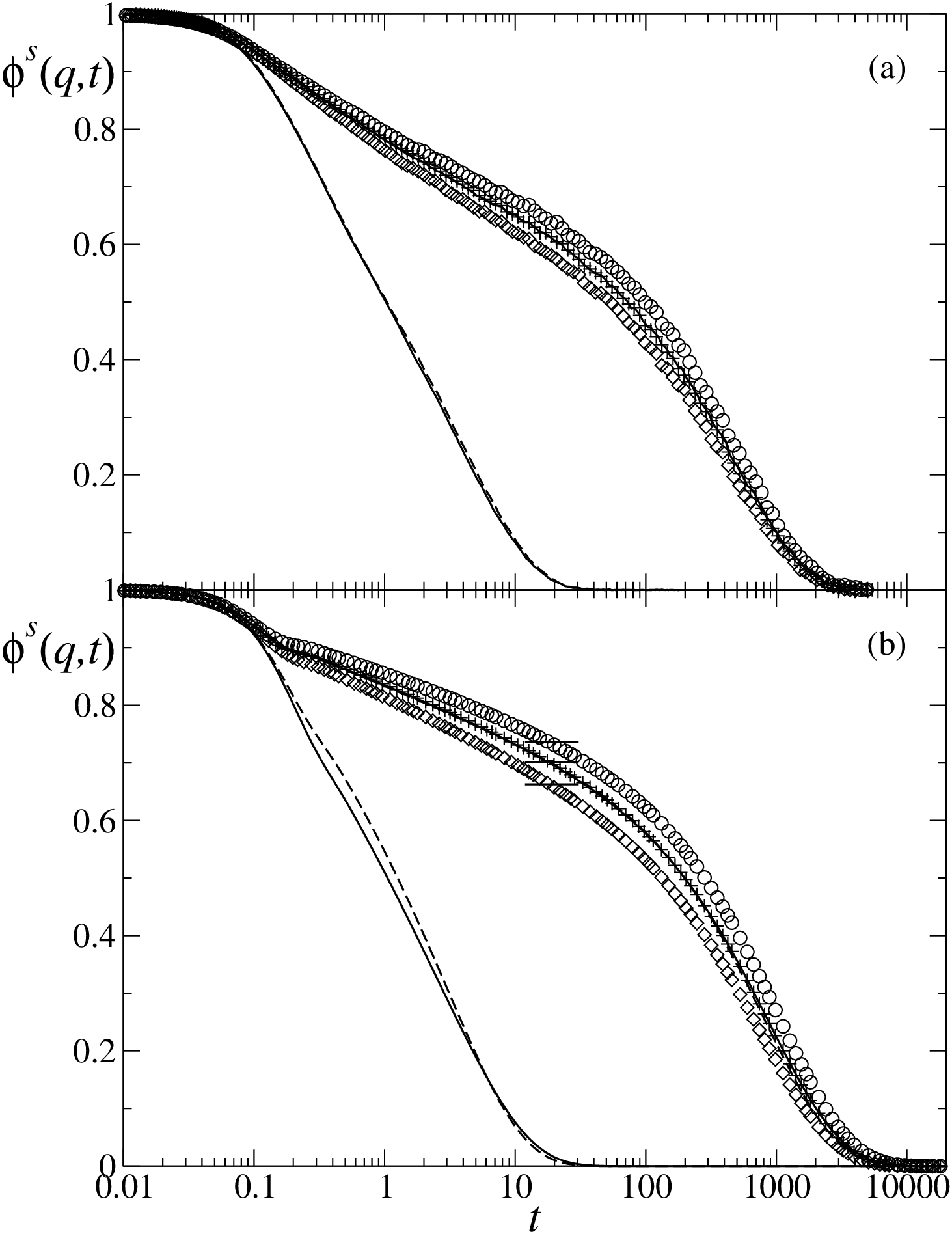}
\caption{\label{three-bin}
  (a) Correlation functions $\phi_\alpha^s(q,t)$ for the BD simulation,
  binned into three particle sizes, $\alpha=\text{small}$ (diamonds),
  $\text{medium}$ (plus symbols), and $\text{large}$ (circles); see text
  for details. The data refer to packing fraction $\varphi=0.58$ and
  $qd=7.8$. The averaged quantity $\phi^s_{\text{pd}}(q,t)$
  analyzed in the discussion in detail is plotted
  as a solid line but coincides with the $\alpha=\text{medium}$ curve on the
  scale of the figure. The solid and dashed lines decaying at shorter
  times are the results for $\varphi=0.54$ from the simulation of the
  polydisperse and the monodisperse system, respectively.
  (b) As in (a), but results from the MCT
  calculations for a three-component mixture at packing fraction
  $\varphi=0.505$ and $qd=7.8$. Again, the averaged $\phi^s_{\text{pd}}(q,t)$
  is included as a solid line and is hidden by the $\alpha=\text{medium}$
  curve. The solid line decaying at shorter times is the averaged result
  from the three-component mixture at $\varphi=0.4$. For comparison,
  one-component results at $\varphi=0.42$ and $\varphi=0.507$ are included
  as dashed lines, the latter being obscured by the polydisperse-averaged
  $\varphi=0.505$ result.
}
\end{figure}

A further comparison to the predictions of the multi-component MCT can
be made by binning the particles of the simulation according to their size
into a different number of bins. Let us demonstrate this for the case of
three bins, $\alpha=\text{small}$ (radii $0.9\le d_{\text{small}}<0.96667$),
$\text{medium}$ ($0.96667\le d_{\text{medium}}<1.03333$),
and $\text{large}$ ($1.03333\le d_{\text{large}}\le 1.1$). The thus
obtained three tagged-particle correlation functions $\phi^s_\alpha(q,t)$
can be compared to the three tagged-particle correlation functions amenable
to the MCT calculation in the three-component system.
Figure~\ref{three-bin}a shows as symbols the results from a three-bin analysis
of the BD simulation data at $\varphi=0.58$ and $q=7.8$.
One notices that the largest particles
show the slowest decay, while the smallest particles decay fastest, as is
intuitively expected. In the $\beta$-relaxation window, one finds an
ordering of the plateau values from small to large particles, where the
smallest particles show the smallest plateau.
Again this follows the expectation that the particles are localized more
tightly the larger they are.
These qualitative features are in agreement with the MCT results for the
three-component mixture, as can be inferred from the symbols
Fig.~\ref{three-bin}b. The ordering of the plateau values
is indicated by the horizontal solid lines which represent the results
for $f_\alpha^{s,c}(q)$. Note that the differences in the
relaxation curves for the three components are more pronounced than in the
binned analysis of the polydisperse simulation, which might be related to the
fact that the particle size distribution in the simulation is continuous.
In the MCT calculation, the alpha-relaxation time of the large particles,
measured by $\phi^s_{\text{large}}(q,\tau)=0.1$, is at the wave number chosen
slower by a factor of about $1.48$ compared to that of the small
particles. This change
is slightly higher than in the simulation, where the same trend
applies with a factor of about $1.25$.

The unbinned correlation function from the $\varphi=0.58$ simulation, averaged
over all particles, is included for in Fig.~\ref{three-bin}a, but on
the scale of the plot, it coincides with the correlation function for the
$\alpha=\text{medium}$ bin. Thus, in a sense, polydispersity effects
`average out' in this quantity. A similar effect applies for the MCT results,
but here, the small difference in the critical packing fraction induced
by the polydispersity leads to a shift in the relaxation time scales close
to the glass transition. Still, the one-component correlator
calculated at a slightly higher packing fraction, $\varphi=0.507$, agrees
with the $\alpha=\text{medium}$ correlator from the three-component mixture
at $\varphi=0.505$ on the scale of the figure. This agreement is not too
surprising, because the MCT parameters quantifying the slow relaxation
features apart from $\varphi^c$ do not change significantly between the
monodisperse and the three-component system, as was mentioned above.
At lower $\varphi$, however, small differences become more apparent. This
can be seen in the three-component $\varphi=0.4$ correlator shown in
Fig.~\ref{three-bin}b as a solid line. It compares well with the result
from a monodisperse calculation (the dashed line in Fig.~\ref{three-bin}b),
but at $\varphi=0.42$; and one notices somewhat different curve shapes.
Again, these polydispersity effects are even smaller in the simulations.
The solid line in Fig.~\ref{three-bin}a shows the simulation result for the
polydisperse system at $\varphi=0.54$, which is compared to the result
from the monodisperse simulation at the same density, shown as a dashed line.
Here, the agreement between the two systems is even better; and note that
we did not have to adjust the packing fractions in this comparison.

Thus it appears that this way of representing the polydisperse system
as a three-component mixture leads to a systematic overestimation of
polydispersity effects. For the binary mixtures studied in
Ref.~\cite{Foffi2004}, it was found that MCT even underestimates the size of
the observed mixing effects. If this applies also to our case, the
over-estimation of polydispersity effects by the three-component approach
would be even stronger.

\section{Conclusion}\label{conclusion}

We have performed Newtonian (ND) and strongly damped Newtonian dynamics
(BD) simulations of a polydisperse quasi-hard-sphere system close to the
glass transition. The wave-vector dependent tagged-particle correlation
functions and the mean-squared displacement curves have been analyzed using
the mode-coupling theory of ideal glass transitions (MCT), in order to
provide a stringent test of the complete theory for a reference case.

To ensure that the simulation data show all the qualitative features
predicted by MCT close to the glass transition, we have analyzed both the
ND and BD data in terms of $\alpha$- and $\beta$-scaling, cf.\
Figs.~\ref{alphamaster} and \ref{beta}. This allows us to identify the
time domain, where one can expect MCT to give a quantitative description
of the data, $t>10$ in our units. In particular, both ND and BD agree at
long times up to a trivial time scale. This universality of the structural
relaxation is predicted by the theory, and fulfilled in great detail by
our simulation data. In particular, the $\beta$-scaling parameters and
those qualitative features of the correction-to-scaling amplitudes we could
test, are independent on the type of short-time dynamics. Similarly, an
analysis of the $\alpha$ relaxation with stretched-exponential fits demonstrates
that the wave-number dependent shape of this relaxation is in agreement with
what one expects from MCT. Other parameters, as for example the critical-decay
power law predicted as an asymptotic MCT feature, cannot be extracted from
the simulation data. The analysis reveals that the
highest density studied in our work, $\varphi=0.59$, shows systematic
deviations from the MCT predictions, and can thus not be explained by the
theory.

The main purpose of this paper is to compare the simulation data
to the full solution of the MCT equations. Leaving aside the small
difference between the simulated polydisperse soft spheres and a true
hard sphere system, this comparison is, in principle, free from any
parameters. Since MCT is an approximate theory, one expects, however,
deviations that can be accounted for by allowing some of the
parameters to vary.

We are able to achieve very good agreement between theory and simulations
if we allow for a smooth mapping of packing fractions, $\varphi\mapsto
\varphi_{\text{MCT}}$, and a similar mapping of wave numbers, $q\mapsto q_{\text{MCT}}$.
The reasons for the needed adjustments are well understood.
First, the critical point for the
(ideal) glass transition calculated within is too low. This is compensated
by mapping $\varphi$. The mapping turns out to be almost linear, and hence
inessential in order to understand the slow relaxation close to the
glass transition as a function of the distance to this transition.
Second, we observe a small mismatch in length scales between the simulation
and the MCT results. This can be accounted for by mapping $q$. The
difference in length scales is typically of the order of $15\%$, and
only about $5\%$ for the localization length estimated from the mean-squared
displacement.
It is possible that these discrepancies are to some extent
due to the slight softness of the simulated system.

Given these parameter mappings,
MCT is able to describe the BD-simulation data over most of the density
range studied quantitatively on a $15\%$ level, as demonstrated in
Figs.~\ref{fit-q7} to \ref{fit-q17}. This extends down even to
relatively short times, $t\approx1$, and over a large range of length
scales, from the nearest-neighbour distance down to $qd\approx20$.
At larger length scales (smaller wave numbers), stronger deviations set in,
which are most pronounced in the long-time diffusive regime of the mean-squared
displacement, but also in the $\beta$-relaxation regime,
cf.\ Figs.~\ref{fit-q4} and \ref{fit-msd}.

One has to keep in mind that the kind of comparison we have performed
here is influenced by three conceptually distinct error sources: (i) deviations
due to the comparison of a slightly polydisperse system in the simulations
to a strictly monodisperse one in the theory; (ii) deviations due to incorrect
structure-factor input to MCT; and (iii) deviations inherent to the MCT
approximation. In order to shed more light on the quality of the MCT
approximation itself, we have tried to disentangle these three error sources.

The influence of polydispersity in the studied system is negligible, as we
have pointed out in Sec.~\ref{polydisp} by comparing to a three-component
and a five-component system mimicking the polydispersity distribution
imposed in the simulations.

On the other hand, the second error source, due to approximations made in
describing the static equilibrium structure, has to be considered carefully.
We have chosen to base most of our discussion on MCT results calculated
from the Percus-Yevick structure factor for the hard-sphere system, because
this is closest to a first-principles calculation. However, if one bases MCT
on the structure factor obtained from the simulation, one can improve
the description of the data. Most prominently, this influences the
prediction of the critical point, which shifts from $\varphi^c_{\text{MCT}}\approx0.516$
to $\varphi^c_{\text{MCT}}\approx0.585$, and thus surprisingly close to the
experimentally determined value. At the same time, many of the predictions
based on the PY structure factor remain quantitatively true, such as the
shape of the $f^c(q)$-versus-$q$ distribution, or the asymptotic shape of
the correlation functions. This finding to some extent justifies our approach
of adjusting the packing fraction $\varphi$ in the PY-based MCT calculations.
In principle, a further error source connected with the static-structure
input is the factorization of three-point static correlations in the MCT
vertices, Eqs.~\eqref{mctv} and \eqref{mctvs}. But we expect this purely
technical approximation to have small influence for our system, as
simulation studies of a binary Lennard-Jones mixture \cite{Sciortino2001}
suggest for systems dominated by hard-core repulsion.

The remaining discrepancies between the simulation results and the MCT
predictions for the hard-sphere system are likely to be the ones giving
information about the quality of the MCT approximation itself. These are:

(i) The wave-vector variation of relaxation times.
This is less pronounced in MCT than it is in the simulations. For
large wave numbers, the BD simulation shows slower relaxation than expected
from the theory, while at small $q$, the relaxation is faster than predicted
by MCT. This indicates an error of the theory in
capturing the length-scale dependence of the dynamics. The error at small $q$
might be
more severe, and is most dramatic when one considers the mean-squared
displacement, i.e.\ the diffusion coefficient. The theory predicts that all
structural relaxation time scales are coupled close to the glass transition.
This implies that the product of the diffusion coefficient and a typical
intermediate-length-scale relaxation time, $D\cdot\tau$, should approach
a constant when $\varphi$ approaches $\varphi^c$. For finite
$\varphi^c-\varphi$, MCT predicts a smooth variation that is in qualitative
agreement with the simulation results, cf.\ Fig.~\ref{dtau}. But the
magnitude of this variation is underestimated by a factor of $2.5$.
This can be viewed as a quantitative error that
has, however, a large impact on the description of $D$ or the relaxation
times at small $q$. We could not test
whether the simulation behaves qualitatively different to the MCT prediction
as $\varphi\to\varphi^c$ due to obvious constraints.
In general, our results show that the improper treatment of time-scale
decoupling within MCT is not peculiar to the diffusion coefficient itself,
but rather sets in smoothly at small $q$ in the $\phi^s(q,t)$.

(ii) At short times the description of the relaxation curves with MCT
is insufficient. This is known since long, but it remains an important
question at how large times deviations can still be seen. In our comparison
of strongly damped Newtonian dynamics, is appears that the tagged-particle
correlators can be fitted quite well even down to $t\approx1$, i.e.\
almost `microscopic' time scales.
But a comparison with undamped Newtonian dynamics simulations reveals that
there, short-time corrections can occur even for relatively large times,
up to $t\approx10$ in our case.
This can provide an explanation for recent observations stating that for a
binary mixture obeying strongly damped colloidal dynamics, the MCT description
extended quantitatively down to surprisingly short times
\cite{Voigtmann2003},
whereas a similar comparison of Newtonian MD data was satisfactory only in
the $\alpha$-relaxation regime \cite{Foffi2004}.

(iii)
At the highest packing fraction analyzed in the present work, more dramatic
deviations between the simulation results and MCT occur. They are most
easily observed as a violation of $\alpha$ scaling, and a different scale
behavior of the corresponding relaxation time. Our simulations hint
towards possibly rare events that induce this behavior. But we have not
been able to establish this within reasonable statistics.

\begin{acknowledgments}
We thank W.~G\"otze for valuable comments on the draft.
Th.V.\ thanks for funding through EPSRC Grant GR/S10377/01, for
partial financial support from the Dr.-Ing.\ Leonhard-Lorenz-Stiftung
of the Technische Universit\"at M\"unchen and of the SFB 563, and the
Universit\"at Konstanz for its hospitality during an earlier stage of the
project.
The financial support for A.M.P.\ is provided by the CICYT under project
MAT~2003-03051-CO3-01, and for M.F.\ by Deutsche Forschungsgemeinschaft,
grant Fu~309/3.
\end{acknowledgments}

\bibliography{mct,add}
\bibliographystyle{apsrev}

\end{document}